\documentclass[prc,amsmath,showpacs,twocolumn,floatfix,unsortedaddress]{revtex4}
\usepackage{graphicx}
\usepackage{dcolumn}
\usepackage[pdftex,dvipsnames,usenames]{color}		
\usepackage{epstopdf}
\usepackage{slashed}
\usepackage{verbatim}
\usepackage{mathtools}
\usepackage{longtable}
\usepackage{amssymb}
\usepackage{bbold}
\usepackage{soul}   

\long\def\Omit#1{}

\allowdisplaybreaks[1]
\newcommand*{\tblref}[1]{Table~\ref{tbl:#1}}

\renewcommand*{\eqref}[1]{(\ref{eq:#1})}
\newcommand*{\eqlab}[1]{\label{eq:#1}}

\newcommand*{\seclab}[1]{\label{sec:#1}}

\newcommand{\ket}[1]{\left|{#1}\right >}

\newcommand{\SP}[3]{\left<{#1}\right|{#2}\left|{#3}\right>}

\newcommand{\beq}{\begin{equation}}
\newcommand{\eeq}{\end{equation}}
\newcommand{\nn}{\nonumber}

\newcommand{\etext}[1]{\mbox{$#1$}}

\def\sphlf{spin-1/2 }
\def\spthlf{spin-3/2 }

\newcommand{\tsp}[2]{\left<\mbox{\small{${#1}$}}\right|T \left|\mbox{\small{${#2}$}}\right>}

\begin{document}

\title{On kinematical constraints in  fermion-antifermion systems}
\author{S.\ Stoica}
\email{s.stoica@kvi.nl}
\affiliation{Kernfysisch Versneller Instituut (KVI), University of Groningen, 9747 AA Groningen, The~Netherlands}
\author{M.F.M.\ Lutz}
\email{M.Lutz@gsi.de}
\affiliation{Gesellschaft f\"{u}r Schwerionenforschung (GSI), Planck Str.~1, 64291 Darmstadt, Germany}
\author{O.\ Scholten}
\email{scholten@kvi.nl}
\affiliation{Kernfysisch Versneller Instituut (KVI), University of Groningen, 9747 AA Groningen, The~Netherlands}

\date{\today}

\begin{abstract}

We consider the scattering of fermions off antifermions with spin $1/2$ and $3/2$.
Starting from helicity partial-wave scattering amplitudes we derive transformations
that eliminate all kinematical constraints. Such amplitudes are expected to satisfy
partial-wave dispersion relations and therefore provide a suitable basis for data analysis and
the construction of effective field theories. Our derivation relies on a decomposition
of the various scattering amplitudes into suitable sets of invariant functions.

\end{abstract}

\pacs{$11.55.-m$, $13.75.Cs$, $ 11.80.-m$}
\maketitle


\section{Introduction}

This work addresses an important aspect arising in the study of fermion-antifermion annihilation. Though it is straight
forward to introduce partial-wave scattering amplitudes in the helicity formalism of Jacob and Wick \cite{Jacob:1959at},
it is a nontrivial task to derive transformations that lead to amplitudes that are kinematically unconstrained. Much work
has gone in the derivation of kinematically unconstraint amplitudes for spin-$1/2$ scattering, see for example
Ref.~\cite{CohenTannoudji:1968}. Such amplitudes have so far not been established for systems involving spin-$1/2$
and $3/2$ fermions. Kinematically unconstraint amplitudes are useful for partial-wave analysis or effective field
theory approaches which consider the consequences of micro causality in terms of partial-wave dispersion-integral
representations~\cite{Chew:1957tf,Nakanishi:1962,Berends:1967vi,Gasparyan:2010xz}. It is the purpose of the present work
to derive such amplitudes by suitable transformations of the helicity partial-wave scattering amplitudes. Our results
will be relevant for the PANDA experiment at FAIR, where protons and antiprotons may be annihilated into systems of spin $3/2$ states.

The technique applied in this work has been used previously in studies of two-body scattering systems with photons,
pions and nucleons~\cite{Chew:1957tf,Ball:1961zza,Barut:1963zz,Hara:1964zza,PhysRev.169.1248,PhysRev.142.1187,PhysRev.170.1606,Scadron:1969rw,CohenTannoudji:1968,Bardeen:1969aw}.
A more recent work addressed the scattering of pseudoscalar off vector particles \cite{Lutz:2003fm}.
A possibly related approach is by Chung and collaborators \cite{PhysRevD.48.1225,Chung:2007nn}. It was applied so far
to mesonic systems only.

In an initial step we decompose the scattering amplitude into invariant functions.
For on-shell kinematics such a decomposition is not unique for systems involving particles with nonzero spin.
The task is to find a basis of invariant functions that are free of kinematical constraints. Such amplitudes are
expected to satisfy a Mandelstam dispersion-integral representation~\cite{Mandelstam:1958xc,Ball:1961zza}. At this
point we distinguish kinematical from dynamical singularities. Kinematical singularities are dependent on the
particularities of the used basis and typically occur at thresholds or pseudo-thresholds. Dynamical singularities
are related to interesting physics such as resonances.

In a second step we consider partial-wave projections of the scattering amplitude. Our goal is to establish
partial-wave amplitudes with convenient analytic properties that justify the use of uncorrelated integral-dispersion
relations. In particular we want to avoid cumbersome constraints relating different partial-wave amplitudes.
Our starting point are the helicity partial-wave amplitudes. It is well known that different helicity partial-wave
amplitudes are correlated at various kinematical conditions. This is readily seen if the helicity partial-wave amplitudes
are expressed in terms of a basis of invariant functions. The kinematical constraints in the helicity partial-wave
amplitudes are eliminated by means of nonunitary transformation matrices that map the initial, respectively final
helicity sates to new covariant states. Since the mapping procedure is based on the exclusive use of on-shell matrix
elements there is no model-dependent off-shell dependence in our considerations.

The work is organized as follows. Section II introduces the conventions used for the kinematics and the spin-$1/2$
and $3/2$ helicity wave functions.
It follows a section where the scattering amplitudes are decomposed into sets of invariant amplitudes.
We consider all two-body reactions possible with spin-$1/2$ and $3/2$ fermions.
In section IV the helicity partial-wave amplitudes are constructed within the given convention. The central results
are presented in section V, where the transformation to partial-wave amplitudes free of kinematical constraints are
derived and discussed.

\section{Wave functions and conventions}

We introduce the 4-momenta $p_{1}$ and $\bar{p}_{1}$ of the incoming and outgoing particles and those of the
anti-particles, $p_{2}$ and $\bar{p}_{2}$. In the center of mass frame we write
\begin{eqnarray}
&& p_{1}^\mu= \left(E_{1},0,0,+p \right)\,, \quad  \bar{p}_{1}^\mu= \left(\bar{E}_1, +\bar{p}\sin{\theta},0 , +\bar{p}\cos\theta \right)\,,
\nonumber\\
&& E_{1}= \sqrt{M_{1}^{2} + p^2}   \,,  \qquad \qquad \bar{E}_{1}= \sqrt{\bar{M}_{1}^{2} + \bar{p}^2}\,,
\nonumber\\
&& p_{2}^\mu= \left(E_{2},0,0,-p \right)\,, \quad  \bar{p}_{2}^\mu= \left(\bar{E}_2, -\bar{p}\sin{\theta},0 , -\bar{p}\cos\theta \right)\,,
\nonumber\\
&& E_{2}= \sqrt{M_{2}^{2} + p^2}   \,,  \qquad \qquad \bar{E}_{2}= \sqrt{\bar{M}_{2}^{2} + \bar{p}^2}\,,
\label{eq:momenta}
\end{eqnarray}
where $\theta$ is the scattering angle, $p$ and $\bar{p}$ are the magnitudes of the initial and final three-momenta.
The relative momenta can be expressed in terms of the total energy $\sqrt{s}$ of the system
\begin{eqnarray}
w^\mu = p_{1}^\mu+p_{2}^\mu = \bar{p}_{1}^\mu + \bar{p}_{2}^\mu\,, \qquad s = w^2\,.
\label{def-w}
\end{eqnarray}
It holds
\begin{eqnarray}
&& p^2=\frac{(s-(M_1 +M_2)^2)(s-(M_1 -M_2)^2)}{4\,s}
\nonumber\\
&& \bar{p}^2=\frac{(s-(\bar{M}_1 +\bar{M}_2)^2)(s-(\bar{M}_1 - \bar{M}_2)^2)}{4\,s}\,.
\label{p-sqr}
\end{eqnarray}
We continue with the specification of helicity eigenstates. This is an important part of the documentation since
there are various distinct phase conventions being used in the literature.
For the spin-$1/2$ fermions the Dirac spinors for the outgoing states are
\begin{eqnarray}
&& u(\bar{p}_{1},\pm \etext{\frac{1}{2}})=
        \left(
        \begin{array}{c}
          \sqrt{\frac{\bar{E}_{1}+\bar{M}_{1}}{2 \bar{M}_{1}}} \,e^{-\frac{i}{2}\,\sigma_y\theta}\, \chi_{\pm}\\
          \pm \sqrt{\frac{\bar{E}_{1}-\bar{M}_{1}}{2 \bar{M}_{1}}}\,e^{-\frac{i}{2}\,\sigma_y\theta}\,\chi_{\pm}\\
        \end{array}
        \right)\,,
\nonumber\\
&& u(\bar{p}_{2},\pm \etext{\frac{1}{2}})=
        \left(
        \begin{array}{c}
          \sqrt{\frac{\bar{E}_{2}+\bar{M}_{2}}{2 \bar{M}_{2}}} \,e^{-\frac{i}{2}\,\sigma_y\theta}\, \chi_{\mp}\\
          \pm \sqrt{\frac{\bar{E}_{2}-\bar{M}_{2}}{2 \bar{M}_{2}}}\,e^{-\frac{i}{2}\,\sigma_y\theta}\,\chi_{\mp}\\
        \end{array}
        \right)\,,
\nonumber\\
&&  \chi_+=\left(%
\begin{array}{c}
  1 \\
  0 \\
\end{array}%
\right)\,, \qquad \quad \chi_-=\left(%
\begin{array}{c}
  0 \\
  1 \\
\end{array}%
\right)  \,.
\label{def-spin12}
\end{eqnarray}
The spinors of the incoming states are recovered by setting $\theta=0$ and removing the bars in (\ref{def-spin12}).
It holds the completeness relation
\begin{eqnarray}
&&   \sum_{\lambda=\pm\frac{1}{2}}\, u(p,\lambda)\, \bar{u}(p,\lambda)
   = {\slashed{p}+ M \over 2\,M} \,.
\end{eqnarray}

We specify the helicity states for a spin-3/2 particle~\cite{Rarita:1941mf}. Most conveniently
they are expressed in terms of spin-one-half and spin-one wave functions
\begin{eqnarray}
&& u^{\mu} (\bar{p}_{1},\etext{\pm \frac{3}{2}) } =
        \epsilon^\mu (\bar{p}_1, \pm 1)\, u(\bar{p}_{1},\pm \etext{\frac{1}{2}})\,,
\label{eq:delta_spinor-1}\\
&&u^{\mu} (\bar{p}_{1},\etext{\pm \frac{1}{2}) } =
        \sqrt{\etext{\frac{2}{3}}}\,\epsilon^{\mu} (\bar{p}_{1}, 0 )\, u(\bar{p}_{1},\pm \etext{\frac{1}{2}})
        \nonumber\\
&& \qquad  \qquad   \quad     + \,\sqrt{\etext{\frac{1}{3}}}\,\epsilon^{\mu} (\bar{p}_{1},\pm 1 ) \,
u(\bar{p}_{1},\mp \etext{\frac{1}{2}})\,,
\nonumber\\
&& \epsilon^\mu(\bar{p}_1, \pm 1)=\left(
                           \begin{array}{c}
                             0 \\
                              \frac{\mp\cos \theta }{\sqrt{2}} \\
                            \frac{- i}{\sqrt{2}} \\
                              \frac{ \pm \sin \theta }{\sqrt{2}} \\
                           \end{array}
                         \right)\,,\;
\epsilon^\mu(\bar{p}_1,  0)=      \left(
                           \begin{array}{c}
                             \frac{\bar p}{\bar M_1} \\
                             \frac{\bar{E}_1}{\bar M_1} \sin \theta \\
                             0 \\
                             \frac{\bar{E}_1}{\bar M_1} \cos \theta \\
                           \end{array}
                         \right)\,,       \nonumber
\end{eqnarray}
and
\begin{eqnarray}
&&u^{\mu} (\bar{p}_2,\etext{\pm \frac{3}{2}) } =
        \epsilon^\mu (\bar{p}_2, \pm 1)\,u(\bar{p}_2,\pm \etext{\frac{1}{2}})
        \label{eq:delta_spinor-2}\\
&&u^{\mu} (\bar{p}_2,\etext{\pm \frac{1}{2}) } =
        \sqrt{\etext{\frac{2}{3}}}\,\epsilon^\mu (\bar{p}_2, 0 )\,u(\bar{p}_2,\pm \etext{\frac{1}{2}})
\nonumber\\
        && \qquad \qquad \quad - \,\sqrt{\etext{\frac{1}{3}}}\,\epsilon^\mu (\bar{p}_2, \pm 1 )\,u(\bar{p}_2,\mp \etext{\frac{1}{2}})\,,
\nonumber\\
&& \epsilon^\mu(\bar{p}_2, \pm 1) = \left(
                           \begin{array}{c}
                             0 \\
                            \frac{\pm \cos \theta }{\sqrt{2}} \\
                            \frac{-i}{\sqrt{2}} \\
                            \frac{\mp \sin \theta }{\sqrt{2}} \\
                           \end{array}
                         \right) \,, \;
\epsilon^\mu(\bar{p}_2, 0) = \left(
                           \begin{array}{c}
                             \frac{\bar p}{\bar M_2} \\
                             -\frac{\bar{E}_2}{\bar M_2} \sin \theta \\
                             0 \\
                             -\frac{\bar{E}_2}{\bar M_2} \cos \theta \\
                           \end{array}
                         \right) \,.\nonumber
\end{eqnarray}
Again the spinor of the corresponding initial states is recovered with $\theta =0$ and by removing the bars
in (\ref{eq:delta_spinor-1}, \ref{eq:delta_spinor-2}). The main properties of the spinors are recalled with
\begin{eqnarray}
&&   \sum_{\lambda=\pm \frac{3}{2},\pm\frac{1}{2}}\, u^\mu(p,\lambda) \,\bar{u}^\nu(p,\lambda)
   = {(\slashed{p}+ M) \over 2\,M} \left[-g ^{\mu\nu}  + \frac{1}{3}\, \gamma^\mu\, \gamma^\nu \right.
\nonumber \\
&& \qquad \quad    \left. +\,\frac{2}{3 \,M^2}\, p^\mu \,p^\nu - \frac{1}{3\, M}
\left( p^\mu \,\gamma^\nu - p^\nu\, \gamma^\mu\right)\right]\,,
\nonumber\\
&&  \gamma^{\mu} \, u_\mu(p, \lambda) = 0\,, \qquad \quad
  p^\mu \, u_\mu(p, \lambda) =0 \;.
\end{eqnarray}

It remains to detail the phase convention of the anti-particles spinors. We use
\begin{eqnarray}
&& \bar{v}(p_1,\lambda)=u^t(p_1,\lambda)\,C \,, \quad   v(\bar{p}_1,\lambda)=C^{-1}
    \bar{u}^t(\bar{p}_1,\lambda)\,,
\nonumber\\
&& \bar{v}(p_2,\lambda)=u^t(p_2,\lambda)\,C \,, \quad   v(\bar{p}_2,\lambda)=C^{-1}
    \bar{u}^t(\bar{p}_2,\lambda)\,,
\nonumber\\ \\
&& \bar{v}_\mu(p_1,\lambda)=u^t_\mu(p_1,\lambda)\,C \,, \quad   v_\mu(\bar{p}_1,\lambda)=C^{-1}
    \bar{u}_\mu^t(\bar{p}_1,\lambda)\,,
\nonumber\\
&& \bar{v}_\mu(p_2,\lambda)=u_\mu^t(p_2,\lambda)\,C \,, \quad   v_\mu(\bar{p}_2,\lambda)=C^{-1}
    \bar{u}_\mu^t(\bar{p}_2,\lambda)\,,    \nonumber
\end{eqnarray}
where $C = i \,\gamma_2 \,\gamma_0$ is the charge-conjugation matrix and $\lambda$ is the helicity of the spinor.
It follows
\begin{eqnarray}
&&    v(\bar{p}_{2},\pm \etext{\frac{1}{2}})=
        \left(
        \begin{array}{c}
          \mp \sqrt{\frac{\bar{E}_{2}-\bar{M}_{2}}{2 \bar{M}_{2}}} \,i \, \sigma_y \,e^{-\frac{i}{2}\,\sigma_y  \theta}\, \chi_{\mp}\\
           \sqrt{\frac{\bar{E}_{2}+\bar{M}_{1}}{2 \bar{M}_{2}}} \,i \, \sigma_y \,e^{-\frac{i}{2}\,\sigma_y \theta}\,\chi_{\mp}\\
        \end{array}
        \right) \;,
\nonumber\\
&& v_{\mu} (\bar{p}_2,\etext{\pm \frac{3}{2}) } =
        \epsilon_\mu^{*} (\bar{p}_2, \pm 1)\, v(\bar{p}_2,\pm \etext{\frac{1}{2}})\,
\nonumber\\
&& v_{\mu} (\bar{p}_2,\etext{\pm \frac{1}{2}) } =
        \sqrt{\etext{\frac{2}{3}}}\,\epsilon_\mu^{*} (\bar{p}_2, 0 )\, v(\bar{p}_2,\pm \etext{\frac{1}{2}})
\nonumber\\
&&\qquad \qquad \quad - \,\sqrt{\etext{\frac{1}{3}}}\,\epsilon_\mu^{*} (\bar{p}_2, \pm 1 )\, v(\bar{p}_2,\mp \etext{\frac{1}{2}}) \,.
\end{eqnarray}

\section{Invariant amplitudes}\seclab{tensors}

We decompose the scattering amplitudes into sets of invariant functions. For this purpose it is convenient to introduce
some further notation
\begin{eqnarray}
&& k={\textstyle{1\over 2}}\,(p_1-p_2)\,, \qquad \bar k
={\textstyle{1\over 2}}\,(\bar p_1-\bar p_2)  \,.
\label{def-k}
\end{eqnarray}
We will consider first the scattering of spin-$1/2$ particles. It is described by the on-shell scattering amplitude
\begin{eqnarray}
&& T_{\frac{1}{2} \bar{ \frac{1}{2}} \to \frac{1}{2} \bar{ \frac{1}{2}}} =
\sum_{i=1}^{4} \sum_\pm F^\pm_i \,\langle \, T^{(i)}_\pm\, \rangle_{\frac{1}{2} \bar{ \frac{1}{2}} \to \frac{1}{2} \bar{ \frac{1}{2}}}
\nonumber\\
&& \qquad \qquad \,
= \sum_{i=1}^{4} \sum_\pm F^\pm_i \,\langle \, T^{(i)}\, \rangle^\pm_{\frac{1}{2} \bar{ \frac{1}{2}} \to \frac{1}{2} \bar{ \frac{1}{2}}}
\,, \eqlab{def-T-1}\\
&& \begin{array}{ll}
T^{(1)}= \mathbb{1} \times \mathbb{1}\,, & \qquad \qquad   T^{(2)}= \gamma_\mu  \times \gamma^\mu \,, \\
T^{(3)}= \mathbb{1} \times \bar{\slashed{k}}\,,  & \qquad \qquad  T^{(4)}= \slashed{k} \times \mathbb{1}\,,\nonumber
\end{array}
\end{eqnarray}
in terms of eight scalar functions $F_i^\pm$. The on-shell particle and antiparticle spinors are implied by the
$\langle \cdots \rangle $ and
$\langle \cdots \rangle^{\pm}$ notations used in \eqref{def-T-1}, \begin{eqnarray}
&& \langle  \bar{\Gamma} \times \Gamma \rangle^\pm_{\frac{1}{2} \bar{ \frac{1}{2}} \to \frac{1}{2} \bar{ \frac{1}{2}}} =
\langle  \bar{\Gamma}  P_\pm \otimes  P_\pm \Gamma \rangle_{\frac{1}{2} \bar{ \frac{1}{2}} \to \frac{1}{2} \bar{ \frac{1}{2}}} =
\nonumber\\ && \qquad
 \Big(\bar{u}(\bar{p}_1)\,\bar{\Gamma}\, P_\pm\,v(\bar{p}_2)\Big)
 \, \Big(\bar{v}(p_2)\,P_\pm\,\Gamma \,u(p_1)\Big)\,,
 \eqlab{def-bracket-1} \\
&& P_+ = \mathbb{1} \,, \qquad \quad P_- = \gamma_5\,, \nonumber
\end{eqnarray}
where for notational clarity the reference to the helicities of initial and final states is not made explicit.

The number of invariant amplitudes needed is determined by the number of helicity amplitudes.
Since we are assuming parity conservation the total number of independent helicity amplitudes is
\begin{eqnarray}
\etext{\frac{1}{2}} \,(2 \,S_{1}+1)\,(2 \,S_{2}+1)\,(2 \,\bar S_{1}+1)\,(2 \,\bar S_{2}+1)\,,
\label{def-number-of-helicity-amplitudes}
\end{eqnarray}
where $S_{1}, S_{2}$ and $\bar S_{1},\bar S_{2}$ are the spins of the initial and final particles.

We assure that our decomposition \eqref{def-T-1} and the ones presented below imply that the invariant amplitudes
$F_i^\pm$ satisfy a Mandelstam dispersion integral representation free of kinematical constraints. They are functions
of the Mandelstam variables $s$, $t$ and $u$.

The particular choice of tensors in \eqref{def-T-1} is not unique. The scattering amplitude considered at off-shell
kinematics requires a more general decomposition. For instance the structures $\slashed{w}\, \gamma_5 \otimes \gamma_5$,
which is not considered in \eqref{def-T-1}, is on-shell equivalent to $(\bar M_1+\bar M_2)\,\gamma_5 \otimes \gamma_5$.

Our bases are characterized by their property that if any structure not part of a basis is decomposed into the basis set
no kinematical singularities arise in the on-shell limit. The identification of such bases sets is a nontrivial task
getting more and more tedious as the spin of the involved particles increase. A good starting point are tensors which
involve minimal powers of momenta.


We continue with the reactions involving a single spin-three-half particle only. There are four
cases to be considered. A decomposition analogous to \eqref{def-T-1} is established. While \eqref{def-T-1}
involves eight scalar functions $F_{1-4}^\pm$ the replacement of a spin-one-half by a spin-three-half particle leads to
16 scalar functions. For notational convenience the invariant amplitudes are denoted with $F_{1-8}^\pm$, even though
this implies a conflict of notation for some amplitudes. Given the local context a misinterpretation is excluded.
We consider first the two spin-three-half production amplitudes
\begin{eqnarray}
&& T_{\frac{1}{2} \bar{ \frac{1}{2}} \to \frac{3}{2} \bar{ \frac{1}{2}}} =
\sum_{i=1}^{8} \sum_\pm F^\pm_i \,\langle \, T^{(i)}_\mu\, \rangle^{\pm,\,\mu}_{\frac{1}{2} \bar{ \frac{1}{2}} \to \frac{3}{2} \bar{ \frac{1}{2}}}\,,
\label{eq:def-bracket-2}\\
&& \langle  \bar{\Gamma} \times\Gamma \rangle^{\pm,\,\mu}_{\frac{1}{2} \bar{ \frac{1}{2}} \to \frac{3}{2} \bar{ \frac{1}{2}}} =
\bar{u}^\mu (\bar{p}_1)\,\bar{\Gamma}\, P_\mp\,v(\bar{p}_2)\,
\bar{v}(p_2)\,P_\pm\,\Gamma \,u(p_1)\,,
\nonumber\\
&& T_{\frac{1}{2} \bar{ \frac{1}{2}} \to \frac{1}{2} \bar{ \frac{3}{2}}} =
\sum_{i=1}^{4} \sum_\pm F^\pm_i \,\langle \, T^{(i)}_\mu\, \rangle^{\pm,\mu}_{\frac{1}{2} \bar{ \frac{1}{2}} \to \frac{1}{2} \bar{ \frac{3}{2}}}\,,
\label{eq:def-bracket-3}\\
&& \langle  \bar{\Gamma} \times\Gamma \rangle^{\pm,\mu}_{\frac{1}{2} \bar{ \frac{1}{2}} \to \frac{1}{2} \bar{ \frac{3}{2}}} =
\bar{u} (\bar{p}_1)\,\bar{\Gamma}\, P_\mp\,v^\mu(\bar{p}_2)\,
\bar{v}(p_2)\,P_\pm\,\Gamma \,u(p_1)\,,
\nonumber\\
&& \begin{array}{ll}
 T^{(1)}_{\mu} =  1 \times \gamma_\mu\,, & \qquad
T^{(2)}_{\mu} = \slashed{k} \times\gamma_\mu\,, \\
T^{(3)}_{\mu} = k_\mu  \times 1 \,, & \qquad
T^{(4)}_{\mu} = w_\mu\,\gamma^\tau \times \gamma_\tau\,, \\
T^{(5)}_{\mu} = w_\mu  \times 1 \,, & \qquad
T^{(6)}_{\mu} = w_\mu  \times\slashed{\bar k}\,\,, \\
T^{(7)}_{\mu} = w_\mu\,\slashed{k}  \times1 \,, & \qquad
T^{(8)}_{\mu} = k_\mu\,\slashed{k} \times1 \,.
\end{array}
\nonumber
\end{eqnarray}

For the inverse reactions we use a convention generated unambiguously by the tensor choice
in (\ref{eq:def-bracket-2}, \ref{eq:def-bracket-3}), where
any tensor $T_{\mu} $ is replaced by a corresponding tensor $\bar T_{\mu}$ generated using the following rule
\begin{equation}
 T^{(i)}_{\mu}  = \bar \Gamma \times \Gamma
\; \to \;
\bar T^{(i)}_{\mu} = \gamma_0 \,\Gamma^\dagger \gamma_0 \times \gamma_0\,
 \bar{\Gamma}^\dagger\,\gamma_0\Big|_{k \leftrightarrow \bar k}\,.
\label{def-generation}
\end{equation}
The decomposition of the inverse reactions is formally analogous to (\ref{eq:def-bracket-2}, \ref{eq:def-bracket-3})
where the role of the tensors
$T_{\mu}$ is taken over by  $\bar T_{\mu}$. The bracket notation is extended naturally with
\begin{eqnarray}
&& \langle  \bar{\Gamma} \times \Gamma \rangle^{\pm,\, \mu}_{\frac{3}{2} \bar{ \frac{1}{2}} \to \frac{1}{2} \bar{ \frac{1}{2}}} =
\bar{u}(\bar{p}_1)\,\bar{\Gamma}\,P_\pm \, v(\bar{p}_2)\,\bar{v}(p_2)\,P_\mp\,\Gamma \,u^\mu(p_1)\,,
 \nonumber\\
&& \langle  \bar{\Gamma} \times \Gamma \rangle^{ \pm,\,\mu}_{\frac{1}{2} \bar{ \frac{3}{2}} \to \frac{1}{2} \bar{ \frac{1}{2}}} =
\bar{u}(\bar{p}_1)\,\bar{\Gamma}\, P_\pm\,v(\bar{p}_2)\,\bar{v}^\mu(p_2)\,P_\mp\,\Gamma \,u(p_1)\,.
\nonumber
\end{eqnarray}


We proceed with reactions involving two spin-three-half particles. There are six different cases to be studied.
It suffices to detail two reactions, the remaining are implied unambiguously within our notations and conventions.
The first four cases are introduced with
\begin{eqnarray}
&& T_{\frac{3}{2} \bar{ \frac{1}{2}} \to \frac{3}{2} \bar{ \frac{1}{2}}} =
\sum_{i=1}^{16} \sum_\pm F^\pm_i \,\langle \, T^{(i)}_{\bar \mu \mu}\, \rangle^{\pm,\bar \mu \mu}_{\frac{3}{2} \bar{ \frac{1}{2}} \to \frac{3}{2} \bar{ \frac{1}{2}}}  \,,
\eqlab{def-bracket-4}\\
&& \langle  \bar{\Gamma} \times \Gamma \rangle^{\pm,\bar \mu \mu}_{\frac{3}{2} \bar{ \frac{1}{2}} \to \frac{3}{2} \bar{ \frac{1}{2}}} =
\Big( \bar{u}^{\bar \mu}(\bar{p}_1)\,\bar{\Gamma} P_\mp v(\bar{p}_2) \Big) \,
 \nonumber\\ && \hspace*{2.5cm} \times\,
\Big( \bar{v}(p_2) P_\mp\Gamma \,u^\mu(p_1) \Big) \,,
\nonumber\\
&& \begin{array}{ll}  \renewcommand{\arraystretch}{1.4}
T^{(1)}_{\bar \mu \mu} = g_{\bar \mu \mu} \, \mathbb{1} \times \mathbb{1} \,,  & \qquad
T^{(2)}_{ \bar \mu \mu} = \gamma_\mu \times \gamma_{\bar \mu}
\\
T^{(3)}_{ \bar \mu \mu} =w_\mu \times \gamma_{\bar \mu} \,,  & \qquad
T^{(4)}_{\bar \mu \mu} = \gamma_\mu \times w_{\bar \mu} \,,
\\
T^{(5)}_{ \bar \mu \mu} =\bar k_\mu \times \gamma_{\bar \mu} \,, & \qquad
T^{(6)}_{ \bar \mu \mu} = \gamma_\mu \times k_{\bar \mu} \,,
\\
T^{(7)}_{\bar \mu \mu} =w_\mu \times w_{\bar \mu} \,, & \qquad
T^{(8)}_{ \bar \mu \mu} = w_\mu \,\gamma_\tau \times \gamma^\tau\,w_{\bar \mu} \,,
\\
T^{(9)}_{ \bar \mu \mu} =w_\mu \times k_{\bar \mu} \,, & \qquad
T^{(10)}_{ \bar \mu \mu} = w_\mu \,\gamma_\tau \times \gamma^\tau\,k_{\bar \mu}
\\
T^{(11)}_{ \bar \mu \mu} = \bar k_\mu \times w_{\bar \mu} \,,& \qquad
T^{(12)}_{ \bar \mu \mu} = \bar k_\mu \,\gamma_\tau \times \gamma^\tau\,w_{\bar \mu} \,,
\\
T^{(13)}_{ \bar \mu \mu} =w_\mu \,\slashed{k}\times w_{\bar \mu} \,, & \qquad
T^{(14)}_{ \bar \mu \mu} = w_\mu \times \slashed{\bar k}\,w_{\bar \mu} \,,
\\
T^{(15)}_{ \bar \mu \mu} = w_\mu \,\slashed{k}\times k_{\bar \mu} \,, & \qquad
T^{(16)}_{ \bar \mu \mu} = \bar k_\mu \times \slashed{\bar k}\,w_{\bar \mu} \,.
\end{array} \nonumber
\end{eqnarray}
Any of the three additional reactions which involve a spin-three-half particle or antiparticle in the initial
and final state is decomposed in terms of the tensors introduced in (\ref{eq:def-bracket-3}, \ref{eq:def-bracket-4}).
The corresponding bracket notation follows from (\ref{eq:def-bracket-3}, \ref{eq:def-bracket-4}) upon moving the Lorentz indices
to the appropriate spinors.


It is left to detail the reactions involving a spin-three-half particle and antiparticle in either
the initial or final state. We write
\begin{eqnarray}
&& T_{\frac{3}{2} \bar{ \frac{3}{2}} \to \frac{1}{2} \bar{ \frac{1}{2}}} =
\sum_{i=1}^{16} \sum_\pm F^\pm_i \,\langle \, T^{(i)}_{\mu \nu}\, \rangle^{\pm,\,\mu \nu}_{\frac{3}{2} \bar{ \frac{3}{2}} \to \frac{1}{2} \bar{ \frac{1}{2}}} \,,
\eqlab{def-bracket-5}\\
&& \langle  \bar{\Gamma} \times \Gamma \rangle^{\pm,\, \mu \nu}_{\frac{3}{2} \bar{ \frac{3}{2}} \to \frac{1}{2} \bar{ \frac{1}{2}}} =
\bar{u}(\bar{p}_1)\,\bar{\Gamma}\,P_\pm\, v(\bar{p}_2)
  \nonumber\\ && \hspace*{2.5cm}
\times\, \bar{v}^\nu(p_2)\,P_\pm\,\Gamma \,u^\mu(p_1)\,,
\nonumber\\
&& \begin{array}{ll}  \renewcommand{\arraystretch}{1.4}
T^{(1)}_{\mu \nu} = g_{\mu \nu}\,\mathbb{1} \times \mathbb{1}    \,,  & \qquad
T^{(2)}_{ \mu \nu} = g_{\mu \nu} \,\gamma_\tau \times \gamma^{\tau} \,,
\\
T^{(3)}_{ \mu \nu} = g_{\mu \nu} \,\slashed{k}\times \mathbb{1} \,, & \qquad
T^{(4)}_{ \mu \nu} = g_{\mu \nu} \,\mathbb{1} \times \slashed{\bar k} \,,
\\
T^{(5)}_{ \mu \nu} = \gamma_\mu \times w_\nu  \,,  & \qquad
T^{(6)}_{ \mu \nu} = \gamma_\mu \times \bar k_\nu \,,
\\
T^{(7)}_{ \mu \nu} = w_\mu \times w_\nu  \,,  & \qquad
T^{(8)}_{ \mu \nu} = w_\mu \,\gamma_\tau \times \gamma^\tau\,w_\nu \,,
\\
T^{(9)}_{ \mu \nu} = \bar k _\mu \times \bar k _\nu  \,,  & \qquad
T^{(10)}_{ \mu \nu} = \bar k_\mu \,\gamma_\tau \times \gamma^\tau\,\bar k_\nu \,,
\\
T^{(11)}_{ \mu \nu} = \bar k _\mu \times w_\nu  \,,  & \qquad
T^{(12)}_{ \mu \nu} = w_\mu \, \gamma_\tau \times \gamma^\tau\,\bar k_\nu \,,
\\
T^{(13)}_{ \mu \nu} =w_\mu \slashed{k} \times w_\nu\,, & \qquad
T^{(14)}_{ \mu \nu} = w_\mu \times \slashed{\bar k}\,w_\nu\,,
\\
T^{(15)}_{ \mu \nu} =\bar k_\mu \times \slashed{\bar k } \,\bar k_{\nu} \,, & \qquad
T^{(16)}_{ \mu \nu} = w_\mu \times \slashed{\bar k } \,\bar k_{\nu}\,,
\end{array} \nonumber
\end{eqnarray}
and refer to (\ref{def-generation}) as a definition of the inverse reaction.


We continue with the reactions involving three spin-three-half particles. There are four cases to be studied.
\begin{eqnarray}
&& T_{\frac{3}{2} \bar{ \frac{1}{2}} \to \frac{3}{2} \bar{ \frac{3}{2}}} =
\sum_{i=1}^{32} \sum_\pm F^\pm_i \,\langle \, T^{(i)}_{\bar \mu \bar \nu \mu}\, \rangle^{\pm,\,\bar \mu \bar \nu \mu}_{\frac{3}{2} \bar{ \frac{1}{2}} \to \frac{3}{2} \bar{ \frac{3}{2}}}\,,
\eqlab{def-bracket-6}\\
&& \langle  \bar{\Gamma} \times \Gamma \rangle^{\pm,\, \bar \mu \bar \nu \mu}_{\frac{3}{2} \bar{ \frac{1}{2}} \to \frac{3}{2} \bar{ \frac{3}{2}}} =
 \Big( \bar{u}^{\bar{\mu}}(\bar{p}_1) \,\bar{\Gamma}\,P_\pm\, v^{\bar{\nu}}(\bar{p}_2) \Big)
  \nonumber\\ && \hspace*{2.5cm}
\times\, \Big( \bar{v}(p_2) \,P_\pm\,\Gamma \,u^\mu(p_1) \Big) \,,
\nonumber\\
&& T_{\frac{1}{2} \bar{ \frac{3}{2}} \to \frac{3}{2} \bar{ \frac{3}{2}}} =
\sum_{i=1}^{32} \sum_\pm F^\pm_i \,\langle \, T^{(i)}_{\bar \mu \bar \nu \mu}\, \rangle^{\pm,\,\bar \mu \bar \nu \mu}_{\frac{1}{2} \bar{ \frac{3}{2}} \to \frac{3}{2} \bar{ \frac{3}{2}}}\,,
\eqlab{def-bracket-7}\\
&& \langle  \bar{\Gamma} \times \Gamma \rangle^{\pm,\, \bar \mu \bar \nu \mu}_{\frac{1}{2} \bar{ \frac{3}{2}} \to \frac{3}{2} \bar{ \frac{3}{2}}} =
 \Big( \bar{u}^{\bar{\mu}}(\bar{p}_1)\,\bar{\Gamma}\,P_\pm\, v^{\bar{\nu}}(\bar{p}_2) \Big)
  \nonumber\\ && \hspace*{2.5cm}
\times\, \Big( \bar{v}^{\mu}(p_2)\,P_\pm\,\Gamma \,u(p_1) \Big) \,,
\nonumber\\
&& \begin{array}{l@{\;}l}
T^{(1)}_{\bar \mu \bar \nu \mu}= g_{\bar{\mu} \bar{\nu}} \, \gamma_{\mu} \times \mathbb{1} \,, &
T^{(2)}_{\bar \mu \bar \nu \mu}= g_{\bar{\mu} \mu} \mathbb{1} \times \gamma_{\bar{\nu}} \,, \\
T^{(3)}_{\bar \mu \bar \nu \mu}= g_{\bar{\mu} \bar{\nu}} \, \mathbb{1} \times w_{\mu} \,, &
T^{(4)}_{\bar \mu \bar \nu \mu}= g_{\bar{\mu} \bar{\nu}} \, \gamma_{\tau} \times \gamma^{\tau} \, w_{\mu}\,, \\
T^{(5)}_{\bar \mu \bar \nu \mu}= g_{\bar{\mu} \mu} \, w_{\bar{\nu}} \times \mathbb{1} \,, &
T^{(6)}_{\bar \mu \bar \nu \mu}= w_{\bar{\mu}} \, \gamma_{\mu} \times \gamma_{\bar{\nu}} \,, \\
T^{(7)}_{\bar \mu \bar \nu \mu}= g_{\bar{\mu} \bar{\nu}} \, \mathbb{1} \times \bar{k}_{\mu} \,, &
T^{(8)}_{\bar \mu \bar \nu \mu}= g_{\bar{\mu} \bar{\nu}} \, \gamma_{\mu} \times \slashed{\bar{k}} \,, \\
T^{(9)}_{\bar \mu \bar \nu \mu}= g_{\bar{\mu} \mu} \, k_{\bar{\nu}} \times \mathbb{1} \,, &
T^{(10)}_{\bar \mu \bar \nu \mu}= g_{\bar{\mu} \mu} \, \slashed{k} \times \gamma_{\bar{\nu}} \,, \\
T^{(11)}_{\bar \mu \bar \nu \mu}= g_{\bar{\mu} \bar{\nu}} \, \slashed{k} \times w_{\mu} \,, &
T^{(12)}_{\bar \mu \bar \nu \mu}= g_{\bar{\mu} \bar{\nu}} \times \slashed{\bar{k}} \, w_{\mu} \,, \\
T^{(13)}_{\bar \mu \bar \nu \mu}= w_{\bar{\nu}} \times \gamma_{\bar{\mu}} \, w_{\mu} \,, &
T^{(14)}_{\bar \mu \bar \nu \mu}= w_{\bar{\mu}} \, w_{\bar{\nu}} \, \gamma_{\mu} \times \mathbb{1} \,, \\
T^{(15)}_{\bar \mu \bar \nu \mu}= g_{\bar{\mu} \bar{\nu}} \, \mathbb{1} \times \slashed{\bar{k}} \, \bar{k}_{\mu} \,, &
T^{(16)}_{\bar \mu \bar \nu \mu}= k_{\bar{\mu}} \times \gamma_{\bar{\nu}} \, w_{\mu} \,, \\
T^{(17)}_{\bar \mu \bar \nu \mu}= g_{\bar{\nu} \mu} \, w_{\bar{\mu}} \, \slashed{k} \, \times \mathbb{1} \,, &
T^{(18)}_{\bar \mu \bar \nu \mu}= w_{\bar{\nu}}  \times \gamma_{\bar{\mu}} \, \bar{k}_{\mu} \,, \\
T^{(19)}_{\bar \mu \bar \nu \mu}= g_{\bar{\mu} \mu} \, k_{\bar{\nu}} \, \slashed{k} \times \mathbb{1} \,, &
T^{(20)}_{\bar \mu \bar \nu \mu}= w_{\bar{\mu}} \, w_{\bar{\nu}} \times w_{\mu} \,, \\
T^{(21)}_{\bar \mu \bar \nu \mu}= w_{\bar{\mu}} \, w_{\bar{\nu}} \, \gamma_{\tau} \times \gamma^{\tau} \, w_{\mu} \,, &
T^{(22)}_{\bar \mu \bar \nu \mu}= k_{\bar{\mu}} \, w_{\bar{\nu}} \times w_{\mu} \,, \\
T^{(23)}_{\bar \mu \bar \nu \mu}= w_{\bar{\mu}} \, w_{\bar{\nu}} \times \bar{k}_{\mu} \,, &
T^{(24)}_{\bar \mu \bar \nu \mu}= k_{\bar{\mu}} \, k_{\bar{\nu}} \times w_{\mu} \,, \\
T^{(25)}_{\bar \mu \bar \nu \mu}= k_{\bar{\mu}} \, k_{\bar{\nu}}  \, \gamma_{\tau} \times \gamma^{\tau} \, w_{\mu} \,, &
T^{(26)}_{\bar \mu \bar \nu \mu}= w_{\bar{\mu}} \, w_{\bar{\nu}} \, \gamma_{\mu} \times \slashed{\bar{k}} \,, \\
T^{(27)}_{\bar \mu \bar \nu \mu}= w_{\bar{\mu}} \, \slashed{k} \times \gamma_{\bar{\nu}} \, w_{\mu} \,, &
T^{(28)}_{\bar \mu \bar \nu \mu}= w_{\bar{\mu}} \, w_{\bar{\nu}} \, \slashed{k} \times w_{\mu} \,, \\
T^{(29)}_{\bar \mu \bar \nu \mu}= w_{\bar{\mu}} \, w_{\bar{\nu}} \times \slashed{\bar{k}} \, w_{\mu} \,, &
T^{(30)}_{\bar \mu \bar \nu \mu}= w_{\bar{\mu}} \, k_{\bar{\nu}} \, \slashed{k} \times w_{\mu} \,, \\
T^{(31)}_{\bar \mu \bar \nu \mu}= w_{\bar{\mu}} \, w_{\bar{\nu}} \times \slashed{\bar{k}} \, \bar{k}_{\mu} \,, &
T^{(32)}_{\bar \mu \bar \nu \mu}= k_{\bar{\mu}} \, k_{\bar{\nu}}  \, \slashed{k} \times w_{\mu} \,.
\end{array} \nonumber
\end{eqnarray}
We use (\ref{def-generation}) to obtain the tensors for the other two inverse reactions.


The scattering of spin-three-half-particles is described by the on shell scattering amplitude
\begin{eqnarray}
&& T_{\frac{3}{2} \bar{ \frac{3}{2}} \to \frac{3}{2} \bar{ \frac{3}{2}}} =
\sum_{i=1}^{64} \sum_\pm F^\pm_i \,\langle \, T^{(i)}_{\bar \mu \bar \nu \mu \nu} \, \rangle^{\pm,\bar \mu \bar \nu \mu \nu}_{\frac{3}{2} \bar{ \frac{3}{2}} \to \frac{3}{2} \bar{ \frac{3}{2}}}\,,
\eqlab{def-bracket-8}\\
&& \langle  \bar{\Gamma} \times \Gamma \rangle^{\pm,\,\bar \mu \bar \nu \mu \nu}_{\frac{3}{2} \bar{ \frac{3}{2}} \to \frac{3}{2} \bar{ \frac{3}{2}}} =
 \Big( \bar{u}^{\bar{\mu}}(\bar{p}_1) \,\bar{\Gamma}\,P_\pm\, v^{\bar{\nu}}(\bar{p}_2) \Big)
  \nonumber\\ && \hspace*{2.5cm}
\times\, \Big( \bar{v}^{\nu}(p_2)\,P_\pm\,\Gamma \,u^\mu(p_1) \Big)\,,
\nonumber
\end{eqnarray}
where the tensors $T^{(i)}_{\bar \mu \bar \nu \mu \nu}$ are given in \tblref{tensors}.

\begin{table*}[htbp]
\renewcommand{\arraystretch}{1.6}
\caption{ Tensor basis for the scattering of spin-three-half particles }
\label{tbl:tensors}
\begin{tabular}{|l l l l|}
\hline
      $T^{(1)}_{\bar \mu \bar \nu \mu \nu}= g_{\bar{\mu} \bar{\nu}} \, \mathbb{1} \times \mathbb{1} \, g_{\mu \nu}$ &
$T^{(2)}_{\bar \mu \bar \nu \mu \nu}= g_{\bar{\mu} \bar{\nu}} \, \gamma_{\tau} \times \gamma^{\tau} \, g_{\mu \nu}$ &
$T^{(3)}_{\bar \mu \bar \nu \mu \nu}= g_{\bar{\nu} \nu}\, g_{\bar{\mu} \mu} \, \mathbb{1} \times \mathbb{1}$ &
$T^{(4)}_{\bar \mu \bar \nu \mu \nu}= g_{\bar{\nu} \nu}\, g_{\bar{\mu} \mu} \, \gamma_{\tau} \times \gamma^{\tau}$ \\
$T^{(5)}_{\bar \mu \bar \nu \mu \nu}= g_{\bar{\mu} \bar{\nu}} \, \slashed{k} \times \mathbb{1} \, g_{\mu \nu}$ &
$T^{(6)}_{\bar \mu \bar \nu \mu \nu}= g_{\bar{\mu} \bar{\nu}} \, \mathbb{1} \times \slashed{\bar{k}} \, g_{\mu \nu}$ &
$T^{(7)}_{\bar \mu \bar \nu \mu \nu}= w_{\bar{\nu}} \times \gamma_{\bar{\mu}} \, g_{\mu \nu}$ &
$T^{(8)}_{\bar \mu \bar \nu \mu \nu}= g_{\bar{\mu} \bar{\nu}} \, \gamma_{\mu} \times w_{\nu}$ \\
$T^{(9)}_{\bar \mu \bar \nu \mu \nu}= k_{\bar{\mu}} \times \gamma_{\bar{\nu}} \, g_{\mu \nu}$ &
$T^{(10)}_{\bar \mu \bar \nu \mu \nu}= g_{\bar{\mu} \bar{\nu}} \, \gamma_{\mu} \times \bar{k}_{\nu}$ &
$T^{(11)}_{\bar \mu \bar \nu \mu \nu}= g_{\bar{\nu} \nu} \, w_{\bar{\mu}} \, \gamma_{\mu} \times \mathbb{1}$ &
$T^{(12)}_{\bar \mu \bar \nu \mu \nu}= g_{\bar{\mu} \nu} \, \mathbb{1} \times \gamma_{\bar{\nu}} \, w_{\mu}$ \\
$T^{(13)}_{\bar \mu \bar \nu \mu \nu}= g_{\bar{\nu} \nu}\, g_{\bar{\mu} \mu} \, \slashed{k} \times \mathbb{1} $ &
$T^{(14)}_{\bar \mu \bar \nu \mu \nu}= g_{\bar{\nu} \nu}\, g_{\bar{\mu} \mu} \, \mathbb{1} \times \slashed{\bar{k}}$ &
$T^{(15)}_{\bar \mu \bar \nu \mu \nu}= w_{\bar{\mu}} \, w_{\bar{\nu}} \times \mathbb{1} \, g_{\mu \nu}$ &
$T^{(16)}_{\bar \mu \bar \nu \mu \nu}= w_{\bar{\mu}} \, w_{\bar{\nu}} \, \gamma_{\tau} \times \gamma^{\tau} \, g_{\mu \nu}$ \\
$T^{(17)}_{\bar \mu \bar \nu \mu \nu}= g_{\bar{\mu} \bar{\nu}} \mathbb{1} \times w_{\mu} \, w_{\nu}$ &
$T^{(18)}_{\bar \mu \bar \nu \mu \nu}= g_{\bar{\mu} \bar{\nu}} \, \gamma_{\tau} \times \gamma^{\tau} \, w_{\mu} \, w_{\nu}$ &
$T^{(19)}_{\bar \mu \bar \nu \mu \nu}= g_{\bar{\mu} \mu} \, w_{\bar{\nu}} \times w_{\nu}$ &
$T^{(20)}_{\bar \mu \bar \nu \mu \nu}= g_{\bar{\nu} \nu} \, w_{\bar{\mu}} \, \gamma_{\tau} \times \gamma^{\tau} \, w_{\mu}$ \\
$T^{(21)}_{\bar \mu \bar \nu \mu \nu}= k_{\bar{\mu}} \, w_{\bar{\nu}} \times \mathbb{1} \, g_{\mu \nu}$ &
$T^{(22)}_{\bar \mu \bar \nu \mu \nu}= g_{\bar{\mu} \bar{\nu}} \, \mathbb{1} \times \bar{k}_{\mu} \, w_{\nu}$ &
$T^{(23)}_{\bar \mu \bar \nu \mu \nu}= k_{\bar{\mu}} \, k_{\bar{\nu}} \times \mathbb{1} \, g_{\mu \nu}$ &
$T^{(24)}_{\bar \mu \bar \nu \mu \nu}= k_{\bar{\mu}} \, k_{\bar{\nu}} \, \gamma_{\tau} \times \gamma^{\tau} \, g_{\mu \nu}$ \\
$T^{(25)}_{\bar \mu \bar \nu \mu \nu}= g_{\bar{\mu} \bar{\nu}} \times \bar{k}_{\mu} \, \bar{k}_{\nu}$ &
$T^{(26)}_{\bar \mu \bar \nu \mu \nu}= g_{\bar{\mu} \bar{\nu}} \, \gamma_{\tau} \times \gamma^{\tau} \, \bar{k}_{\mu} \, \bar{k}_{\nu}$ &
$T^{(27)}_{\bar \mu \bar \nu \mu \nu}= g_{\bar{\mu} \bar{\nu}} \, \gamma_{\nu} \times \slashed{\bar{k}} \, w_{\mu}$ &
$T^{(28)}_{\bar \mu \bar \nu \mu \nu}= g_{\bar{\mu} \mu} \, k_{\bar{\nu}} \times w_{\nu}$ \\
$T^{(29)}_{\bar \mu \bar \nu \mu \nu}= g_{\bar{\mu} \mu} \, w_{\bar{\nu}} \times \bar{k}_{\nu}$ &
$T^{(30)}_{\bar \mu \bar \nu \mu \nu}= w_{\bar{\mu}} \, \slashed{k} \times \gamma_{\bar{\nu}} \, g_{\mu \nu}$ &
$T^{(31)}_{\bar \mu \bar \nu \mu \nu}= g_{\bar{\mu} \nu} \, k_{\bar{\nu}} \, \gamma_{\tau} \times \gamma^{\tau} \, w_{\mu}$ &
$T^{(32)}_{\bar \mu \bar \nu \mu \nu}= g_{\bar{\nu} \nu} \, w_{\bar{\mu}} \, \gamma_{\tau} \times \gamma^{\tau} \, \bar{k}_{\mu}$ \\
$T^{(33)}_{\bar \mu \bar \nu \mu \nu}= w_{\bar{\mu}} \, w_{\bar{\nu}} \, \slashed{k} \times \mathbb{1} \, g_{\mu \nu}$ &
$T^{(34)}_{\bar \mu \bar \nu \mu \nu}= w_{\bar{\mu}} \, w_{\bar{\nu}} \times \slashed{\bar{k}} \, g_{\mu \nu}$ &
$T^{(35)}_{\bar \mu \bar \nu \mu \nu}= g_{\bar{\mu} \bar{\nu}} \, \slashed{k} \times w_{\mu} \, w_{\nu}$ &
$T^{(36)}_{\bar \mu \bar \nu \mu \nu}= g_{\bar{\mu} \bar{\nu}} \, \mathbb{1} \times \slashed{\bar{k}} \, w_{\mu} \, w_{\nu}$ \\
$T^{(37)}_{\bar \mu \bar \nu \mu \nu}= w_{\bar{\mu}} \, w_{\bar{\nu}} \, \gamma_{\mu} \times w_{\nu}$ &
$T^{(38)}_{\bar \mu \bar \nu \mu \nu}= w_{\bar{\nu}} \times \gamma_{\bar{\mu}} \, w_{\mu} \, w_{\nu}$ &
$T^{(39)}_{\bar \mu \bar \nu \mu \nu}= w_{\bar{\mu}} \, k_{\bar{\nu}} \, \slashed{k} \times g_{\mu \nu}$ &
$T^{(40)}_{\bar \mu \bar \nu \mu \nu}= g_{\bar{\mu} \bar{\nu}} \, \mathbb{1} \times \slashed{\bar{k}} \, w_{\mu} \, \bar{k}_{\nu}$ \\
$T^{(41)}_{\bar \mu \bar \nu \mu \nu}= g_{\bar{\nu} \mu} \, w_{\bar{\mu}} \, \slashed{k} \times w_{\nu}$ &
$T^{(42)}_{\bar \mu \bar \nu \mu \nu}= k_{\bar{\mu}} \times \gamma_{\bar{\nu}} \, w_{\mu} \, w_{\nu}$ &
$T^{(43)}_{\bar \mu \bar \nu \mu \nu}= w_{\bar{\mu}} \, w_{\bar{\nu}} \, \gamma_{\mu} \times \bar{k}_{\nu}$ &
$T^{(44)}_{\bar \mu \bar \nu \mu \nu}= g_{\bar{\mu} \nu} \, w_{\bar{\nu}} \times \slashed{\bar{k}} \, w_{\mu}$ \\
$T^{(45)}_{\bar \mu \bar \nu \mu \nu}= k_{\bar{\mu}} \, k_{\bar{\nu}} \, \slashed{k} \times \mathbb{1} \, g_{\mu \nu}$ &
$T^{(46)}_{\bar \mu \bar \nu \mu \nu}= g_{\bar{\mu} \bar{\nu}} \, \mathbb{1} \times \slashed{\bar{k}} \, \bar{k}_{\mu} \, \bar{k}_{\nu}$ &
$T^{(47)}_{\bar \mu \bar \nu \mu \nu}= k_{\bar{\mu}} \, k_{\bar{\nu}} \, \gamma_{\mu} \times w_{\nu}$ &
$T^{(48)}_{\bar \mu \bar \nu \mu \nu}= w_{\bar{\nu}} \times \gamma_{\bar{\mu}} \, \bar{k}_{\mu} \, \bar{k}_{\nu}$ \\
$T^{(49)}_{\bar \mu \bar \nu \mu \nu}= w_{\bar{\mu}} \, w_{\bar{\nu}} \times w_{\mu} \, w_{\nu}$ &
$T^{(50)}_{\bar \mu \bar \nu \mu \nu}= w_{\bar{\mu}} \, w_{\bar{\nu}} \, \gamma_{\tau} \times \gamma^{\tau} \, w_{\mu} \, w_{\nu}$ &
$T^{(51)}_{\bar \mu \bar \nu \mu \nu}= k_{\bar{\mu}} \, k_{\bar{\nu}} \times w_{\mu} \, w_{\nu}$ &
$T^{(52)}_{\bar \mu \bar \nu \mu \nu}= k_{\bar{\mu}} \, k_{\bar{\nu}} \, \gamma_{\tau} \times \gamma^{\tau} \, w_{\mu} \, w_{\nu}$ \\
$T^{(53)}_{\bar \mu \bar \nu \mu \nu}= w_{\bar{\mu}} \, w_{\bar{\nu}} \times \bar{k}_{\mu} \, \bar{k}_{\nu}$ &
$T^{(54)}_{\bar \mu \bar \nu \mu \nu}= w_{\bar{\mu}} \, w_{\bar{\nu}} \, \gamma_{\tau} \times \gamma^{\tau} \, \bar{k}_{\mu} \, \bar{k}_{\nu}$ &
$T^{(55)}_{\bar \mu \bar \nu \mu \nu}= k_{\bar{\mu}} \, w_{\bar{\nu}} \times w_{\mu} \, w_{\nu}$ &
$T^{(56)}_{\bar \mu \bar \nu \mu \nu}= w_{\bar{\mu}} \, w_{\bar{\nu}} \times \bar{k}_{\mu} \, w_{\nu}$ \\
$T^{(57)}_{\bar \mu \bar \nu \mu \nu}= w_{\bar{\mu}} \, \slashed{k} \times \gamma_{\bar{\nu}} \, w_{\mu} \, w_{\nu}$ &
$T^{(58)}_{\bar \mu \bar \nu \mu \nu}= w_{\bar{\mu}} \, w_{\bar{\nu}} \, \gamma_{\nu} \times \slashed{\bar{k}} \, w_{\mu}$ &
$T^{(59)}_{\bar \mu \bar \nu \mu \nu}= w_{\bar{\mu}} \, w_{\bar{\nu}} \, \slashed{k} \times w_{\mu} \, w_{\nu}$ &
$T^{(60)}_{\bar \mu \bar \nu \mu \nu}= w_{\bar{\mu}} \, w_{\bar{\nu}} \times \slashed{\bar{k}} \, w_{\mu} \, w_{\nu}$ \\
$T^{(61)}_{\bar \mu \bar \nu \mu \nu}= w_{\bar{\mu}} \, k_{\bar{\nu}} \, \slashed{k} \times w_{\mu} \, w_{\nu}$ &
$T^{(62)}_{\bar \mu \bar \nu \mu \nu}= w_{\bar{\mu}} \, w_{\bar{\nu}} \times \slashed{\bar{k}} \, w_{\mu} \, \bar{k}_{\nu}$ &
$T^{(63)}_{\bar \mu \bar \nu \mu \nu}= k_{\bar{\mu}} \, k_{\bar{\nu}} \, \slashed{k} \times w_{\mu} \, w_{\nu}$ &
$T^{(64)}_{\bar \mu \bar \nu \mu \nu}= w_{\bar{\mu}} \, w_{\bar{\nu}} \times \slashed{\bar{k}} \, \bar{k}_{\mu} \, \bar{k}_{\nu}$ \\
\hline
\end{tabular}
\end{table*}

\section{Partial-wave decomposition}
\seclab{partwave}

The helicity matrix elements of the scattering operator, $T$, are decomposed into
partial-wave amplitudes characterized by the total angular momentum $J$.
Given a specific process together with our convention of the helicity wave functions it suffices
to specify the helicity projection of the initial and final wave functions $\lambda_{1},\lambda_2$ and
$\bar \lambda_{1}, \bar \lambda_2$.
We write
\begin{eqnarray}
&& \SP{\bar{\lambda}_1\bar{\lambda}_2}{T}{\lambda_1\lambda_2}= \sum_{J}
(2\, J + \!1) \,\langle \bar{\lambda}_1 \bar{\lambda}_2 | \,T_J | \lambda_1 \lambda_2 \rangle \,
d^{(J)}_{\lambda,\bar{\lambda}} (\theta)\,,
\nonumber\\
&& d^{(J)}_{\lambda, \bar \lambda} (\theta )=(-)^{\lambda-\bar \lambda}\, d^{(J)}_{-\lambda, -\bar \lambda} (\theta )
\nonumber\\
&& \qquad \quad \;= (-)^{\lambda-\bar \lambda}\, d^{(J)}_{\bar \lambda, \lambda} (\theta ) =
d^{(J)}_{-\bar \lambda, -\lambda} (\theta )\,,
\eqlab{def-Wigner} \\
&& \langle \bar{\lambda}_1 \bar{\lambda}_2 | \,T_J | \lambda_1 \lambda_2 \rangle
= \int_{-1}^{-1} \frac{\mathrm{d}\cos\theta}{2} \tsp{\bar{\lambda}_1\bar{\lambda}_2}{\lambda_1 \lambda_2} \,
d^{(J)}_{\lambda,\bar{\lambda}} (\theta)\,, \nonumber
\end{eqnarray}
with $\lambda=\lambda_{1}-\lambda_{2}$ and $\bar{\lambda}=\bar{\lambda}_{1}-\bar{\lambda}_{2}$.
Wigner's rotation functions, $d^{(J)}_{\lambda,\bar \lambda}(\theta)$, are used in a convention as characterized
by \eqref{def-Wigner}.

It is useful to introduce parity eigenstates of good total angular momentum $J$,
formed in terms of the helicity states $\ket{\lambda_{1},\lambda_{2}}_J$ \cite{Jacob:1959at}.
The phase conventions assumed in this work imply the relation
\begin{eqnarray}
&& \langle -\bar{\lambda}_1 -\bar{\lambda}_2 | \,T\, | -\lambda_1 -\lambda_2 \rangle =
\nonumber\\
&& = (-)^{S_{1}-S_{2} +\bar S_1 -\bar S_2 + \lambda-\bar \lambda} \,
\langle \bar{\lambda}_1 \bar{\lambda}_2 | \,T\, | \lambda_1 \lambda_2 \rangle\,.
\label{def-helicity-flip}
\end{eqnarray}
The various parity eigenstates are readily identified. For the
$\frac{1}{2} \bar{ \frac{1}{2}} $ system we introduce
\begin{eqnarray}
 \nn
    \ket{1_{\pm},J}&=& \frac{1}{\sqrt{2}} \left(
        \ket{\etext{+\frac{1}{2}},\etext{ +\frac{1}{2}}}_J \pm
        \ket{\etext{-\frac{1}{2}},\etext{ -\frac{1}{2}}}_J\right)\,,\\
    \ket{2_{\pm},J}&=& \frac{1}{\sqrt{2}} \left(
        \ket{\etext{+\frac{1}{2}},\etext{ -\frac{1}{2}}}_J \pm
        \ket{\etext{-\frac{1}{2}},\etext{ +\frac{1}{2}}}_J\right) \;.
        \eqlab{11-hel-basis}
\end{eqnarray}
For the $\frac{3}{2} \bar{ \frac{1}{2}} $ system,
\begin{eqnarray}
 \nn
    \ket{1_{\pm},J}&=& \frac{1}{\sqrt{2}} \left(
        \ket{\etext{+\frac{1}{2}},\etext{ +\frac{1}{2}}}_J \mp
        \ket{\etext{-\frac{1}{2}},\etext{ - \frac{1}{2}}}_J\right)\,,\\
 \nn
    \ket{2_{\pm},J}&=& \frac{1}{\sqrt{2}} \left(
        \ket{\etext{+\frac{1}{2}},\etext{ -\frac{1}{2}}}_J \mp
        \ket{\etext{-\frac{1}{2}},\etext{ +\frac{1}{2}}}_J\right)\,,\\
 \nn
    \ket{3_{\pm},J}&=&\frac{1}{\sqrt{2}}\left(
        \ket{\etext{+\frac{3}{2}},\etext{+\frac{1}{2}}}_J\mp
        \ket{\etext{-\frac{3}{2}},\etext{-\frac{1}{2}}}_J\right)\,,\\
    \ket{4_{\pm},J}&=&\frac{1}{\sqrt{2}}\left(
        \ket{\etext{+\frac{3}{2}},\etext{-\frac{1}{2}}}_J\mp
        \ket{\etext{-\frac{3}{2}},\etext{+\frac{1}{2}}}_J\right) \;. \eqlab{31-hel-basis}
\end{eqnarray}
For the $ \frac{1}{2} \bar{ \frac{3}{2}} $ system,
\begin{eqnarray}
 \nn
    \ket{1_{\pm},J}&=& \frac{1}{\sqrt{2}} \left(
        \ket{\etext{+\frac{1}{2}},\etext{ +\frac{1}{2}}}_J \mp
        \ket{\etext{-\frac{1}{2}},\etext{ - \frac{1}{2}}}_J\right)\,,\\
 \nn
    \ket{2_{\pm},J}&=& \frac{1}{\sqrt{2}} \left(
        \ket{\etext{+\frac{1}{2}},\etext{ -\frac{1}{2}}}_J \mp
        \ket{\etext{-\frac{1}{2}},\etext{ +\frac{1}{2}}}_J\right)\,,\\
 \nn
    \ket{3_{\pm},J}&=&\frac{1}{\sqrt{2}}\left(
        \ket{\etext{-\frac{1}{2}},\etext{-\frac{3}{2}}}_J\mp
        \ket{\etext{+\frac{1}{2}},\etext{+\frac{3}{2}}}_J\right)\,,\\
    \ket{4_{\pm},J}&=&\frac{1}{\sqrt{2}}\left(
        \ket{\etext{+\frac{1}{2}},\etext{-\frac{3}{2}}}_J\mp
        \ket{\etext{-\frac{1}{2}},\etext{+\frac{3}{2}}}_J\right) \;. \eqlab{13-hel-basis}
\end{eqnarray}
For the $\frac{3}{2} \bar{ \frac{3}{2}} $ system,
\begin{eqnarray}
 \nn
    \ket{1_{\pm},J}&=& \frac{1}{\sqrt{2}} \left(
            \ket{\etext{+\frac{1}{2}},\etext{ + \frac{1}{2}}}_J \pm
            \ket{\etext{-\frac{1}{2}},\etext{ - \frac{1}{2}}}_J\right)\,,\\
 \nn
    \ket{2_{\pm},J}&=&\frac{1}{\sqrt{2}}\left(
        \ket{\etext{+\frac{3}{2}},\etext{+\frac{3}{2}}}_J\pm
        \ket{\etext{-\frac{3}{2}},\etext{-\frac{3}{2}}}_J\right)\,,\\
 \nn
    \ket{3_{\pm},J}&=& \frac{1}{\sqrt{2}} \left(
            \ket{\etext{+\frac{1}{2}},\etext{- \frac{1}{2}}}_J \pm
            \ket{\etext{-\frac{1}{2}},\etext{+ \frac{1}{2}}}_J\right)\,,\\
 \nn
    \ket{4_{\pm},J}&=&\frac{1}{\sqrt{2}}\left(
            \ket{\etext{+\frac{3}{2}},\etext{+\frac{1}{2}}}_J\pm
            \ket{\etext{-\frac{3}{2}},\etext{-\frac{1}{2}}}_J\right)\,,\\
 \nn
    \ket{5_{\pm},J}&=&\frac{1}{\sqrt{2}}\left(
        \ket{\etext{-\frac{1}{2}},\etext{-\frac{3}{2}}}_J\pm
        \ket{\etext{+\frac{1}{2}},\etext{+\frac{3}{2}}}_J\right)\,,\\
 \nn
    \ket{6_{\pm},J}&=&\frac{1}{\sqrt{2}}\left(
        \ket{\etext{+\frac{1}{2}},\etext{-\frac{3}{2}}}_J\pm
        \ket{\etext{-\frac{1}{2}},\etext{+\frac{3}{2}}}_J\right)\,,\\
 \nn
    \ket{7_{\pm},J}&=&\frac{1}{\sqrt{2}}\left(
        \ket{\etext{+\frac{3}{2}},\etext{-\frac{1}{2}}}_J\pm
        \ket{\etext{-\frac{3}{2}},\etext{+\frac{1}{2}}}_J\right)\,,\\
    \ket{8_{\pm},J}&=&\frac{1}{\sqrt{2}}\left(
        \ket{\etext{+\frac{3}{2}},\etext{-\frac{3}{2}}}_J\pm
        \ket{\etext{-\frac{3}{2}},\etext{+\frac{3}{2}}}_J\right) \;. \eqlab{33-hel-basis}
\end{eqnarray}
The partial-wave helicity amplitudes $t^{J}_{\pm,ij}$ that carry good angular momentum $J$ and good parity are defined with
\begin{eqnarray}
t^{J}_{\pm,ij} =  \SP{i_{\pm},J\,}{T}{j_{\pm},J}\,,
\eqlab{def-tij}
\end{eqnarray}
where $i$ and $j$ label the states. The unitarity condition takes the simple form
\begin{eqnarray}
\Im \Big[t_\pm^{J} (s)\Big]^{-1}_{ij} &=& - \frac{M_1\,M_2}{2\,\pi}\frac{p_i}{\sqrt{s}}\,\delta_{ij} \, ,
\label{def-unitarity}
\end{eqnarray}
where $M_1$ and $M_2$ are the masses of the particles of the intermediate state.

\section{Covariant partial-wave projectors}

From a field theoretical point of view a two-body scattering amplitude is determined by the Bethe-Salpeter equation
in terms of an interaction kernel defined by two-particle irreducible Feynman diagrams and a fully dressed two-particle
propagator. To establish a connection we consider first the limit of short-range forces.
Correspondingly, we consider quasi-local two-body interaction terms as they arise naturally in any type of effective
field theory. It is almost obvious that for such structures the Bethe-Salpeter equation can be solved by algebraic
methods (see e.g. \cite{Nieves:1998hp,Lutz:1999yr,Nieves:1999bx,Lutz:2001mi,Lutz:2001dr,Lutz:2001yb,Lutz:2003fm}).

The off-shell scattering amplitude
\begin{eqnarray}
&&T(\bar k,k,w) = \sum_{J,\pm,a,b} \,T^{J}_{\pm,ab}(s)\,{\cal Y}^{J}_{\pm,ab}(\bar k, k,w)
\nonumber\\
&& \qquad \qquad \;\; + \,T^{\rm off-shell}(\bar k,k,w)
\label{def-on-shell}
\end{eqnarray}
can be decomposed into on-shell partial-wave amplitudes  $T^{J}_{\pm,ab}(s)$ and a set of
projectors $Y^{J}_{\pm,ab}(\bar k, k,w) $ that carry well defined total angular momentum
$J$ and parity \cite{Lutz:1999yr,Lutz:2001mi,Lutz:2001yb,Lutz:2003fm}. The indices $a,b$ reflect the possible states with given
total angular momentum and parity (see (\ref{eq:11-hel-basis}-\ref{eq:33-hel-basis})). The
remainder $T^{\rm off-shell}(\bar k,k,w)$ vanishes for on-shell kinematics.
While the  partial-wave amplitudes  $T^{J}_{\pm,ab}(s)$ depend on the total energy $s$ only, the projectors
are fully off-shell quantities. The latter are nothing but suitably constructed polynomials in
the 4-momenta $\bar k_\mu, k_\mu, w_\mu,$ and $ \gamma_\mu$. Depending on the reaction considered they
may carry open Lorentz indices. By construction a partial-wave projector ${\cal Y}^{J}_{\pm,ab}(\bar k, k,w) $
is non-vanishing only for its associated angular momentum $J$ and parity in the center of mass frame.
As a consequence the projectors are a convenient tool to solve the Bethe-Salpeter equation in the limit
of short-range forces.

In the absence of spin the construction of the projectors ${\cal Y}^{J}_{\pm}$ is trivial. It is implied by
the identification
\begin{eqnarray}
&& Y_J =  \left(\frac{\bar{r}^2 \,r^2 }{s^2}\right)^{J/2}
P_J\left(-\frac{\bar{r} \cdot r}{\sqrt{\bar{r}^2 \,r^2}}\right)\,,
\label{projectors-scalar}
\end{eqnarray}
where we have introduced the notation
\begin{eqnarray}
\bar{r}^\mu &=& \bar{k}^\mu - \frac{\bar k \cdot w}{s}w^\mu\;, \nonumber\\
{r}^\mu &=& {k}^\mu-\frac{ k \cdot w}{s}w^\mu \;,
\label{def-r}
\eqlab{r}
\end{eqnarray}
with the Legendre polynomials $P_J$. For a given angular momentum  $J$ the projector
as given in (\ref{projectors-scalar}) is a polynomial in the three off-shell momenta $\bar r, r,w$ after
multiplication with the factor $s^J$. The latter reflects a particular normalization of the projectors that
imply their associated phase-space function to be asymptotically bounded.

While in the scalar case the covariant-partial wave amplitudes $T^{J}_\pm (s)$  are obtained from the
helicity-partial wave amplitudes $t^{J}_\pm(s)$ by dividing out the phase-space factor
$ \left(\bar p\,p /s \right)^J$, for the cases of interest in this work the analogous relations
are significantly more complicated. They take the form
\begin{align}
T^{J}_\pm (s) &= \left( \frac{s}{  \bar{p} \,p} \right)^J \big[\bar{U}_{\pm}^{J}(s) \big]^T\, t^J_\pm(s) \, U^J_{\pm}(s) \;,
  \eqlab{MfromUsandT}
\end{align}
with nontrivial matrices $U_{\pm}(s)$ and $\bar{U}_{\pm}(s)$ characterizing the transformation
for the initial and final states from the helicity basis to the new kinematic-free basis. This implies a change
in the phase-space distribution:
\begin{align}
    \nn
    \rho_{\pm}^{J}(s ) &= - \Im\Big[T_{\pm}^{J}(s)\Big]^{-1} \\
        & = \frac{M_1 \, M_2 }{2 \,\pi} \left(\frac{p}{\sqrt{s}}\right)^{2\,J+1}
    \Big[ U^J_{\pm}(s)\Big]^{-1} \Big[U_{\pm}^{J}(s)\Big]^{T,-1}\,.
    \label{def-rho}
\end{align}
Like in the scalar case we adapt a convention for the transformation matrices that lead to an asymptotically bounded
phase-space matrix, i.e. we require
\begin{eqnarray}
\lim_{s\to \infty}\det \rho _{\pm}^{J}( s ) =  {\rm const} \neq 0\,.
\label{def-rho-constraint}
\end{eqnarray}
It is important to realize that all covariant partial wave amplitudes are subject to kinematical
constraints at $s=0$. Dividing out powers of s will not introduce additional kinematical singularities,
but it will just change their realization.

The significance of introducing the kinematic-free basis lies in the identification and elimination of kinematical
constraints. Helicity-partial-wave amplitudes are correlated at specific kinematical conditions.
This is seen once the amplitudes $t^{J}_{\pm,ij}(\sqrt{s})$ are expressed in terms of the invariant functions
$ F_{i}^{\pm}(s,t)$. If one ignores such correlations it is impossible to reconstruct the invariant
functions from the partial-wave amplitudes in an unambiguous manner. This is a well know problem related
to the use of the helicity basis in  co-variant models, see for example the review Ref.~\cite{CohenTannoudji:1968}.
In contrast covariant-partial wave amplitudes $T^{J}_\pm(s) $ are free of kinematical constraints and can therefore
be used  efficiently in partial-wave dispersion relation.

In the following we will present the required transformation matrices relevant for this work. To the best knowledge
of the authors they are novel and not presented in the literature before. We provide a particularly detailed presentation
for the $ \frac{1}{2} \bar{ \frac{1}{2}} \to \frac{1}{2} \bar{ \frac{1}{2}} $ case, but refrain from
giving the tedious details for the remaining cases.

Any scattering matrix can be expressed in terms of the tensor set \eqref{def-T-1}. To construct the kinematic-free
basis it is therefore a necessary condition that the matrix elements of the tensors \eqref{def-T-1} evaluated in
the new basis are free from singularities. After some calculations this condition leads to the transformation:
\begin{eqnarray}
&& U^J_{\pm,\frac{1}{2} \bar{ \frac{1}{2}}}= \left(
\begin{array}{rr}
 \frac{\sqrt{2 \,J+1}\,\sqrt{s}}{\sqrt{2}\, \alpha_{\mp}} & 0
 \\
 \mp\frac{\sqrt{J\,(2\, J+1)}\,\beta_{\pm}\,\sqrt{s}}{\sqrt{2\,(J+1)} \alpha_{\pm} \,\alpha_{\mp}} &\frac{\sqrt{2\, J+1} \,p}{\sqrt{2 J\,(J+1)} \alpha_{\pm}}
\end{array}
\right)\; ,
\nonumber\\
&& \bar{\alpha}_\pm =
    \sqrt{\frac{\bar{E}_1+\bar{M}_1}{2\, \bar{M}_1}}\sqrt{\frac{\bar{E}_2\pm \bar{M}_2}{2\,\bar{M}_2}}
\nonumber\\
&& \quad \; \; +\,\sqrt{\frac{\bar{E}_1-\bar{M}_1}{2\, \bar{M}_1}} \sqrt{\frac{\bar{E}_2\mp \bar{M}_2}{2 \,\bar{M}_2}} \; ,
\nonumber\\
&&  \bar{\beta}_\pm = \sqrt{\frac{\bar{E}_1+\bar{M}_1}{2\, \bar{M}_1}}\sqrt{\frac{\bar{E}_2\pm \bar{M}_2}{2\, \bar{M}_2}}
\nonumber\\
&& \quad \;\;    -\,\sqrt{\frac{\bar{E}_1-\bar{M}_1}{2\, \bar{M}_1}} \sqrt{\frac{\bar{E}_2\mp \bar{M}_2}{2\, \bar{M}_2}} \,.
\eqlab{trafo-NbarN}
\end{eqnarray}
The expression for $\alpha_\pm$ and $\beta_\pm $ follow from the definition above by removing the bars.

The phase-space distribution becomes:
\begin{eqnarray}
&&\rho_{\pm,\frac{1}{2} \bar{ \frac{1}{2}}}^{\,J}(s) = \frac{1}{(2\, J+1) \,\pi} \left(\frac{p}{\sqrt{s}}\right)^{2\,J-1}
\nonumber\\
&& \quad    \, \times    \left(
        \begin{array}{cc}
    \frac{p^2\,\left(s-M_{\pm}^2\right)}{4 s^2}&\frac{J\,p^2\,M_{\pm}}{2\,s\,\sqrt{s}}\\
    \frac{J\,p^2\,M_\pm}{2\,s\,\sqrt{s}} & \frac{J\,(s-M_{\mp}^2)\,\left((J+1)\,s + J\,M_\pm^2\right)}{4 s^2}
      \end{array}
    \right),
    \label{rho12}
\end{eqnarray}
where $M_\pm = M_1 \pm M_2$. We observe the consistency of the phase-space distribution (\ref{rho12})
with the desired asymptotic behavior (\ref{def-rho-constraint}).

The construction of the associated projectors involves the polynomials introduced already
in (\ref{projectors-scalar}) and their derivatives defined as
\begin{eqnarray}
&&  Y'_J=  \left(\frac{\bar{r}^2 \,r^2}{s^2}\right)^{(J-1)/2}\,
P'_J\left(-\frac{\bar{r} \cdot r}{\sqrt{\bar{r}^2 \,r^2}}\right)\,,
\nonumber\\
&& Y''_J =  \left(\frac{\bar{r}^2 \,r^2}{s^2}\right)^{(J-2)/2}
P''_J\left(-\frac{\bar{r} \cdot r}{\sqrt{\bar{r}^2 \,r^2}}\right)\;,
\label{projectors-scalar-der}
\end{eqnarray}
obeying identities that can easily be derived from those of the Legendre polynomials such as
\begin{eqnarray}
&& (2\,J+1)\, Y_{J} = Y'_{J+1} - \frac{\bar{r}^2 \,r^2}{s^2}\, Y'_{J-1}\,,
\nonumber \\
&& (J+1) \,Y_{J+1} = - \frac{\bar{r}\cdot  r}{s}\, (2\,J+1)\,Y_{J} - \frac{\bar{r}^2\, r^2}{s^2}\,J\,Y_{J-1} \,,
\nonumber \\
&& (J-n) \,Y^{(n)}_{J} = - \frac{\bar r \cdot r}{s}\, Y^{(n+1)}_{J} - \frac{\bar{r}^2\, r^2}{s^2}\,Y^{(n+1)}_{J-1}\,,
\end{eqnarray}
where $n$ denotes the order of the derivative. The projectors read
\begin{eqnarray}
&& {\cal Y}_{\pm,11}^{J} =  \frac{\mp 1}{s}  \, \Big( P_\pm \otimes  P_\pm\Big)\,Y_J \,,
\nonumber\\
&& {\cal Y}_{\pm,12}^{J}
	 =  \frac{\mp 1}{s}  \, \Big( P_\pm \otimes P_\pm \, \tilde \gamma_{\nu }\Big)\,\Big[
\frac{\bar{r}^\nu}{\sqrt{s}} \, Y_J' + \frac{r^\nu}{\sqrt{s}} \,\frac{\bar{r}^2}{s} \, Y_{J-1}'\Big]\,,
\nonumber\\
&& {\cal Y}_{\pm,21}^{J}
	=  \frac{ \mp 1}{s} \,\Big( \tilde \gamma_{\mu }\,P_\pm \otimes P_\pm \Big)\,\Big[
\frac{r^\mu}{\sqrt{s}} \, Y_J' + \frac{\bar{r}^\mu}{\sqrt{s}} \,\frac{r^2}{s} \, Y_{J-1}' \Big]\,,
\nonumber
\\
&& {\cal Y}_{\pm,22}^{J} = \frac{\pm 1}{s} \, \Big( \tilde \gamma_{\mu }\,P_\pm \otimes P_\pm\, \tilde \gamma_{\nu } \Big)    \,
 \Big[ g^{\mu\nu} Y_J' + 2\,\frac{\bar{r}^\mu r^\nu} {s}  \, Y_{J-1}'
\nonumber\\
&& \qquad     	-
\frac{\bar{r}^\nu}{\sqrt{s}} \, \Big(
\frac{r^\mu}{\sqrt{s}} \, Y_J'' + \frac{\bar{r}^\mu}{\sqrt{s}} \,\frac{r^2}{s} \, Y_{J-1}'' \Big)
\nonumber\\
&& \qquad  - \frac{r^\nu}{\sqrt{s}} \,\frac{\bar{r}^2}{s} \, \Big(
\frac{r^\mu}{\sqrt{s}} \, Y_{J-1}'' + \frac{\bar{r}^\mu}{\sqrt{s}} \,\frac{r^2}{s} \, Y_{J-2}'' \Big)\ \Big]  \,,
\label{result-projectors}
\end{eqnarray}
where $\tilde{\gamma}^\mu={\gamma}^\mu- \slashed{w}\, w^\mu/s$.
The important merit of (\ref{result-projectors}) lies in its regularity in the 4-momenta.
This implies that its associated partial-wave amplitudes are suitable for use in dispersion-integral equations.

The partial-wave amplitudes can be computed in terms of the invariant amplitudes $F_i^\pm(s,t)$. The
expressions require an average over $ x =\cos{\theta} $ of the center of mass frame.  We established the
following result
\begin{widetext}
  \begin{align}
  \label{eq:MJ-NNNN}
&T^{J}_{\pm,11} = \mp \frac{(2\,J+1)}{2}\,\Big[ \,
       s\,A^{J}_{\pm,1} + \bar{M}_{\mp}\,M_{\mp}\,\left(A^{J}_{\pm,2} +  \frac{\bar{M}_{\pm}}{2}\,A^{J}_{\pm,3} + \frac{M_{\pm}}{2}\,A^{J}_{\pm,4} \right)
       + \frac{J\,\bar{M}_{\pm}\,M_{\pm}}{(J+1)}\,A^{J}_{\mp,2}
\\ \nonumber &
       - \frac{(2\,J+1)\,\bar{M}_{\pm}\,M_{\pm}\,\left(\bar{M}_{\mp}^2 - s\right)\,\left(M_{\mp}^2 - s\right)}{4\,(J+1)\,s^2}\,A^{J+1}_{\pm,2}
       - \frac{M_{\pm}\,\bar{p}^2\,\left(M_{\mp}^2 - s\right)}{2\,s}\,A^{J+1}_{\pm,3}
       - \frac{\bar{M}_{\pm}\,p^2\,\left(\bar{M}_{\mp}^2 - s\right)}{2\,s} \,A^{J+1}_{\pm,4}
      \Big]\,,
\\
\nonumber &
T^{J}_{\pm,12} = \mp \frac{s^{3/2}}{2} \,A^{J-1}_{\pm,3} \mp \frac{(2\,J+1)\,\bar{M}_{\pm}}{4\,(J+1)\sqrt{s}}\Big[\left(M_{\pm}^2 - s\right)\,A^{J}_{\mp,2} - \frac{p^2\,\left(\bar{M}_{\mp}^2-s\right)}{s} \,A^{J+1}_{\pm,2}\Big]
    \pm \frac{\bar{p}^2\,p^2}{2\,\sqrt{s}}\,A^{J+1}_{\pm,3}\,,
\\
\nonumber &
T^{J}_{\pm,21} = \mp \frac{s^{3/2}}{2} \,A^{J-1}_{\pm,4} \mp
     \frac{(2\,J+1)\,M_{\pm}}{4\,(J+1)\sqrt{s}}\Big[\left(\bar{M}_{\pm}^2 - s\right)\,A^{J}_{\mp,2} - \frac{\bar{p}^2\,\left(M_{\mp}^2-s\right)}{s} \,A^{J+1}_{\pm,2}\Big] \pm \frac{\bar{p}^2\,p^2}{2\,\sqrt{s}}\,A^{J+1}_{\pm,4}\,,
\\
\nonumber &
T^{J}_{\pm,22} = \pm \frac{s}{2\,J}\,A^{J-1}_{\pm,2}
      \mp \frac{(2\,J+1)\,\left(\bar{M}_{\pm}^2 - s\right)\,\left(M_{\pm}^2 - s\right)}{8\,\left(J + J^2\right)\,s} \,
       A^{J}_{\mp,2}
      \pm \frac{\bar{p}^2\,p^2\,A^{J+1}_{\pm,2}}{2\,(J+1)\,s}\,,
\end{align}
\end{widetext}
where $\bar M_{\pm } =\bar M_1 \pm \bar M_2$ and $ M_{\pm } = M_1 \pm  M_2$ and
\begin{eqnarray}
&& A^{J}_{\pm,i}(s) = \left( \frac{s}{\bar p\,p} \right)^J \int_{-1}^{1} \!
\mathrm{d} \cos\theta \, F_i^\pm(s,t) \, P_J(\cos \theta) \,,\nonumber
\nonumber\\
&& F_i^\pm(s,t) \Big|_{\small \mbox{on shell}}=\sum_J \frac{2\,J+1 }{2} A^J_{\pm,i}(s) \, Y_J\,,
\end{eqnarray}
with the Legendre polynomials $P_J(\cos \theta)$. The expressions (\ref{eq:MJ-NNNN}) were
obtained in application of the general results of Appendix A.
From our expression for $T^{J}_{\pm,ab}$ and ${\cal Y}^J_{\pm, ab}$ it follows that there are no kinematical
constraints left. Any correlation in the partial-wave amplitudes would manifest itself in a singular behavior
in the expressions for the projectors. Conversely, a correlation in the projectors would lead to a singular
structure in the expressions for $T^{J}_{\pm,ab}$.  All the coefficients in front of the invariant
amplitudes $A^{J}_{\pm,i}(s)$ are regular confirming our claim.

Though the presentation given so far proves our claim, it leaves obscure how to arrive at our result.
The transformation $U_\pm^J$ can be derived by a polynomial ansatz for the invariant function $F_i$, as it arises
in the limit of short-range forces.  As an initial condition the expressions for the partial-wave amplitudes
$T_\pm^J$ are made regular. This condition by itself is not sufficient since it may lead to projectors that are
singular or equivalently to amplitudes $T_\pm^J$ that are still correlated. The absence of such residual correlations
is most efficiently proven by the construction of their associated projectors.

As an example for the construction of the projectors we elaborate the derivation of ${\cal Y}_{\pm,21}^J(\bar k, k,w)$.
Given \eqref{trafo-NbarN}, the contribution of each of the invariant amplitudes \eqref{def-T-1} to the $21$
partial-wave amplitude can be calculated. The various contributions multiplied by their associated on-shell
tensors define
a scattering amplitude which constitutes an on-shell version of the projector ${\cal Y}^{J}_{\pm, ab}(\bar k, k,w)$.
To be specific we obtain for the on-shell version of the projector associated with $T^J_{\pm, 21}$,
\begin{eqnarray}
&&s^{\frac{3}{2}}\, {\cal Y}_{\pm,21}^J \Big|_{\rm on-shell}  =
\mp Y_J'\,T_4^\pm   \pm   \bar M_\mp \frac{M_+  M_-}{2\,s}\,Y_J'\,T_1^\pm
\nonumber\\
&& \qquad \pm \,T_1^\pm  \,Y_{J-1}' \, \frac{\bar M_\pm}{2\,s^2} \, (s-\bar M_{\mp}^2)\, p^2\,,
\eqlab{on-shell-projector}
\end{eqnarray}
expressed in terms of the on-shell tensor basis \eqref{def-T-1}.
An important property of the  projectors is their independence on any mass parameter. Any mass
factor $M_i$ or $\bar M_i$ occurring in \eqref{on-shell-projector} can be eliminated in favor of
an appropriately chosen factor $\slashed{p_i}$ or $\slashed{\bar p_i}$. This leads to the
following identities,
\begin{eqnarray}
&& \langle \bar r^\mu \,\Big( \tilde \gamma_{\mu }\,P_\pm \otimes P_\pm \Big) \rangle
  =\left( \bar M_\pm/2 -\frac{\bar{k} \cdot w}{s}\, \bar M_\mp \right) \langle T_1^\pm \rangle
\nonumber\\
&& \qquad  =\frac{\bar M_\pm}{2\,s} \left(s -\bar M_{\mp}^2 \right)  \,\langle T_1^\pm  \rangle\;,
\nonumber \\
&& \langle r^\mu \,\Big( \tilde \gamma_{\mu }\,P_\pm \otimes P_\pm \Big)\rangle
  = \langle T_4^\pm \rangle -\frac{k \cdot w}{s} \,\bar M_\mp\, \langle T_1^\pm\rangle
\nonumber\\
&& \qquad  = \langle T_4^\pm \rangle-\frac{M_+ M_-}{2\,s} \,\bar M_\mp \,\langle T_1^\pm\rangle \;,
\label{res-44}
\end{eqnarray}
where we used \eqref{r}. Note also $(E_1-E_2)\sqrt{s}=2\,(w\cdot k)=M_+ M_-$ and $p^2=-r^2$. Applying (\ref{res-44})
to \eqref{on-shell-projector} leads to the result given in (\ref{result-projectors}).

\begin{widetext}

\begin{table*}[b]
\renewcommand{\arraystretch}{1.9}
\begin{tabular}{|l|l|}
\hline
 $ (1,1) : \frac{\sqrt{3} s}{{p} {\alpha}_{\pm}}$  &
 $ (2,1) :  \pm \sqrt{\frac{3 J}{J+1}}\frac{M_{\mp} \sqrt{s}}{{p} {\alpha}_{\pm}}$\\
\hline
 $ (2,2) :  \mp \sqrt{\frac{3 J}{J+1}}\frac{\sqrt{s}}{{\alpha}_{\mp}}$ &
 $ (3,1) : -\sqrt{\frac{J}{J+1}} \left(\frac{2  M_{\pm}^2}{{\alpha}_{\mp}  M_1}+\frac{ M_\mp \sqrt{s}}{{\alpha}_{\pm}
    p} \right) $\\
\hline
 $ (3,2) :  -\sqrt{\frac{J}{J+1}} \frac{{M}_{\pm}^2}{{\alpha}_{\mp}\sqrt{s}}$ &
 $ (3,3) : \sqrt{\frac{J}{J+1}} \frac{ {p}}{{\alpha}_{\pm}}$ \\
\hline
 $ (4,1) : \pm  \sqrt{\frac{J(J-1)}{(J+1)(J+2)}} \left(\frac{2 M_{\pm} s-8  M_1   p^2}{ {\alpha}_{\mp}
    M_1 \sqrt{s}}-\frac{ M_1 s-4  M_\mp   p^2}{{\alpha}_{\pm}  M_1   p}\right)\qquad \qquad \quad \;\;$
 &
 $ (4,2) : \pm \sqrt{\frac{J(J-1)}{(J+1)(J+2)}} \frac{2 \left( {\beta}_{\mp}
 M_1  p-{\alpha}_{\pm} \left(2  p ^2- E_1  \sqrt{s}\right)\right)}{{\alpha}_{\mp} {\alpha}_{\pm}  M_1} \qquad \quad \;\;$\\
\hline
 $ (4,3) : \pm \sqrt{\frac{J(J-1)}{(J+1)(J+2)}} \frac{{p}  M_{\mp}}{{\alpha}_{\pm} \sqrt{s}}$
 &
 $ (4,4) : \pm \sqrt{\frac{J(J-1)}{(J+1)(J+2)}} \frac{{p}^2}{{\alpha}_{\mp}\sqrt{s}}$ \\
\hline
\end{tabular}
\caption{Non zero elements of the transformation matrix for the $\frac{3}{2} \bar{\frac{1}{2}}$ channel.}
\label{tbl:DNtransf}
\vskip0.6cm
\begin{tabular}{|l|l|}
\hline
 $ (1,1) : \frac{\sqrt{3} s}{{p} \,{\alpha}_{\pm}}$
 & $ (2,1) : \pm \sqrt{\frac{3 J}{J+1}}\frac{M_{\mp} \sqrt{s}}{{p} {\alpha}_{\pm}} $\\
 $ (2,1) : \mp \sqrt{\frac{3 J}{J+1}}\frac{\sqrt{s}}{{\alpha}_{\mp}} $ &
 $(3,1) : \sqrt{\frac{J}{J+1}} \left( \mp \frac{2  M_{\pm}^2}{{\alpha}_{\mp}  M_2}+\frac{ M_\mp \sqrt{s}}{{\alpha}_{\pm}  p} \right)$\\
 $(3,2) : \sqrt{\frac{J}{J+1}} \frac{{M}_{\pm}^2}{{\alpha}_{\mp} \sqrt{s}}$
 & $ (3,3) : \mp \sqrt{\frac{J}{J+1}} \frac{ {p}}{{\alpha}_{\pm}} $ \\
 $ (4,1) : \sqrt{\frac{J(J-1)}{(J+1)(J+2)}} \left(\frac{\pm 2  M_{\pm} s-8  M_2   p^2}{ {\alpha}_{\mp}
    M_2 \sqrt{s}}-\frac{ M_2 s \pm 4  M_\mp   p^2}{{\alpha}_{\pm}  M_2   p}\right) \qquad \qquad \quad \;\;$
 &
 $ (4,2) : \sqrt{\frac{J(J-1)}{(J+1)(J+2)}} \frac{ 2\left({\beta}_{\mp}
 M_2  p \pm {\alpha}_{\pm} \left(2  p ^2- E_2  \sqrt{s}\right)\right)}{{\alpha}_{\mp} {\alpha}_{\pm}  M_2} \;\;\; \qquad \quad \;\;$\\
 $ (4,3) : \pm \sqrt{\frac{J(J-1)}{(J+1)(J+2)}} \frac{{p}  M_{\mp}}{ {\alpha}_{\pm} \sqrt{s}}$
 & $ (4,4) : \pm \sqrt{\frac{J(J-1)}{(J+1)(J+2)}} \frac{{p}^2}{{\alpha}_{\mp}\sqrt{s}}$ \\
\hline
\end{tabular}
\caption{Non zero elements of the transformation matrix for the $\frac{1}{2} \bar{\frac{3}{2}}$ channel.}
\label{tbl:NDtransf}
\vskip0.6cm
\begin{tabular}{|l|l|}
\hline
$(1,1) : \frac{3\,s^{3/2}}{\alpha_{\mp}\,p^2}$
&
$(2,1) : -\frac{\left(3\,\alpha_{\mp}^2 + \alpha_{\pm}^2\right)\,s^{3/2}}{\alpha_{\mp}\,p^2}$
\\ \hline
$(2,2) : \frac{4\,\alpha_{\pm}\,s}{p}$
&
$(3,1) : \mp \frac{3\,\beta_{\pm}\,\sqrt{J}\,s^{3/2}}{\alpha_{\mp}\,\alpha_{\pm}\,\sqrt{J + 1}\,p^2}$
\\ \hline
$(3,3) : \frac{3\,s}{\alpha_{\pm}\,\sqrt{J}\,\sqrt{J + 1}\,p}$
&
$(4,1) : -\frac{\sqrt{3}\,\sqrt{J}\,\left(\alpha_{\pm}\,E_{1} + \alpha_{\mp}\,p\right)\,s^{3/2}}{\alpha_{\mp}\,\alpha_{\pm}\,\sqrt{J + 1}\,M_{1}\,p^2}$
\\ \hline
$(4,3) : \mp \frac{\sqrt{3}\,s}{\alpha_{\pm}\,\sqrt{J}\,\sqrt{J + 1}\,p}$
&
$(4,4) : \pm \frac{2\,\sqrt{3}\,s}{\alpha_{\mp}\,\sqrt{J}\,\sqrt{J + 1}\,M_{1}}$
\\ \hline
$(5,1) : \pm \frac{\sqrt{3}\,\sqrt{J}\,\left(\alpha_{\pm}\,E_{2} + \alpha_{\mp}\,p\right)\,s^{3/2}}{\alpha_{\mp}\,\alpha_{\pm}\,\sqrt{J + 1}\,M_{2}\,p^2}$
&
$(5,2) : \mp \frac{4\,\sqrt{3}\,\beta_{\pm}\,\sqrt{J}\,s}{\sqrt{J + 1}\,p}$
\\ \hline
$(5,3) : -\frac{\sqrt{3}\,\left(\alpha_{\mp}^2 + 3\,\alpha_{\pm}^2\right)\,s}{\alpha_{\pm}\,\sqrt{J}\,\sqrt{J + 1}\,p}$
&
$(5,4) : \pm \frac{2\,\sqrt{3}\,s}{\alpha_{\mp}\,\sqrt{J}\,\sqrt{J + 1}\,M_{1}}$
\\ \hline
$(5,5) : \frac{4\,\sqrt{3}\,\alpha_{\mp}\,\sqrt{s}}{\sqrt{J}\,\sqrt{J + 1}}$
&
$(6,1) : \frac{\sqrt{3}\,\sqrt{J - 1}\,\sqrt{J}\,\left(\mp 2\,\beta_{\pm}\,E_{2} + \alpha_{\pm}\,M_{2}\right)\,s^{3/2}}{\alpha_{\mp}\,\alpha_{\pm}\,\sqrt{J + 1}\,\sqrt{J + 2}\,M_{2}\,p^2}$
\\ \hline
$(6,2) : \pm \frac{4\,\sqrt{3}\,\beta_{\pm}^2\,\sqrt{J - 1}\,\sqrt{J}\,s}{\alpha_{\pm}\,\sqrt{J + 1}\,\sqrt{J + 2}\,p}$
&
$(6,3) : \frac{2\,\sqrt{3}\,\left(\alpha_{\mp}\,\beta_{\mp} + 3\,\alpha_{\pm}\,\beta_{\pm}\right)\,\sqrt{J - 1}\,s}{\alpha_{\pm}\,\sqrt{J}\,\sqrt{J + 1}\,\sqrt{J + 2}\,p}$
\\ \hline
$(6,4) : \mp \frac{2\,\sqrt{3}\,\beta_{\pm}\,\sqrt{J - 1}\,s}{\alpha_{\mp}\,\alpha_{\pm}\,\sqrt{J}\,\sqrt{J + 1}\,
      \sqrt{J + 2}\,M_{1}}$
&
$(6,5) : \frac{-4\,\sqrt{3}\,\alpha_{\mp}\,\beta_{\pm}\,\sqrt{J - 1}\,\sqrt{s}}{\alpha_{\pm}\,\sqrt{J}\,\sqrt{J + 1}\,\sqrt{J + 2}}$
\\ \hline
$(6,6) : \frac{2\,\sqrt{3}\,\sqrt{J - 1}\,p\,\sqrt{s}}{\alpha_{\pm}\,\sqrt{J}\,\sqrt{J + 1}\,
      \sqrt{J + 2}\,M_{2}}$
&
$(7,1) : \pm \frac{\sqrt{3}\,\sqrt{J - 1}\,\sqrt{J}\,\left(\beta_{\pm}\,E_{1} - \beta_{\mp}\,p\right)\,s^{3/2}}{\alpha_{\mp}\,\alpha_{\pm}\,\sqrt{J + 1}\,\sqrt{J + 2}\,M_{1}\,p^2}$
\\ \hline
$(7,2) : \mp \frac{4\,\sqrt{3}\,\sqrt{J - 1}\,\sqrt{J}\,p\,s}{\alpha_{\pm}\,\sqrt{J + 1}\,\sqrt{J + 2}\,M_{1}^2}$
&
$(7,3) : \frac{-2\,\sqrt{3}\,\sqrt{J - 1}\,\left(E_{1} \pm 4\,\alpha_{\mp}\,\alpha_{\pm}\,p\right)\,s}{\alpha_{\pm}\,\sqrt{J}\,\sqrt{J + 1}\,\sqrt{J + 2}\,M_{1}\,p}$
\\ \hline
$(7,4) : \frac{-2\,\sqrt{3}\,\left(\pm 4\,\alpha_{\mp}\,\alpha_{\pm}\,\beta_{\mp} + \beta_{\pm}\right)\,\sqrt{J - 1}\,s}{\alpha_{\mp}\,\alpha_{\pm}\,\sqrt{J}\,\sqrt{J + 1}\,\sqrt{J + 2}\,M_{1}}$
&
$(7,5) : \pm \frac{4\,\sqrt{3}\,\alpha_{\mp}\,\sqrt{J - 1}\,\left(\alpha_{\pm}\,E_{1} + \alpha_{\mp}\,p\right)\,\sqrt{s}}{\alpha_{\pm}\,\sqrt{J}\,\sqrt{J + 1}\,\sqrt{J + 2}\,M_{1}}$
\\ \hline
$(7,6) : \pm \frac{2\,\sqrt{3}\,\sqrt{J - 1}\,p\,\sqrt{s}}{\alpha_{\pm}\,\sqrt{J}\,\sqrt{J + 1}\,
      \sqrt{J + 2}\,M_{2}}$
&
$(7,7) : \frac{8\,\sqrt{3}\,\alpha_{\pm}\,\sqrt{J - 1}\,p}{\sqrt{J}\,\sqrt{J + 1}\,
      \sqrt{J + 2}}$
\\ \hline
$(8,1) : \pm \frac{\beta_{\pm}\,\left(3\,\beta_{\mp}^2 + \beta_{\pm}^2\right)\,\sqrt{J - 2}\,\sqrt{J - 1}\,\sqrt{J}\,s^{3/2}}{\alpha_{\mp}\,\alpha_{\pm}\,\sqrt{J + 1}\,\sqrt{J + 2}\,\sqrt{J + 3}\,p^2}$
&
$(8,2) : \mp \frac{2\,\sqrt{J - 2}\,\sqrt{J - 1}\,\sqrt{J}\,\left(\beta_{\pm}^3\,M_{1}^2\,M_{2} \pm
       3\,\beta_{\pm}\,M_{2}\,p^2 + 2\,\alpha_{\mp}\,p^3\right)\,s}{\alpha_{\pm}^2\,\sqrt{J + 1}\,
      \sqrt{J + 2}\,\sqrt{J + 3}\,M_{1}^2\,M_{2}}$
\\ \hline
$(8,3) : \mp \frac{2 \sqrt{J - 2}\,\sqrt{J - 1}\,\alpha_{\mp}\,\left(3\,\left(3 \pm 4\,\alpha_{\mp}^2\right)\,M_{1}^2\,M_{2} +
        16\,\alpha_{\mp}^2\,M_{1}\,p^2 + 12\,M_{2}\,p^2\right)\,\sqrt{s}}{\sqrt{J}\,\sqrt{J + 1}\,
       \sqrt{J + 2}\,\sqrt{J + 3}\,M_{1}\,p^2}$
&
$(8,4) : \frac{16\,\alpha_{\mp}\,\beta_{\mp}\,\sqrt{J - 2}\,\sqrt{J - 1}\,\left(- 3\,\beta_{\pm}\,M_{2} \mp  2\,\alpha_{\mp}\,p\right)\,
      \sqrt{s}}{\sqrt{J}\,\sqrt{J + 1}\,\sqrt{J + 2}\,\sqrt{J + 3}\,p}$
\\ \hline
$(8,5) : \pm \frac{4\,\alpha_{\mp}\,\sqrt{J - 2}\,\sqrt{J - 1}\,\left(\mp 8\,\alpha_{\mp}\,\beta_{\mp}\,M_{2} +
       \left(\alpha_{\mp}^2 + 3\,\alpha_{\pm}^2\right)\,\sqrt{s}\right)}{\sqrt{J}\,\sqrt{J + 1}\,\sqrt{J + 2}\,
      \sqrt{J + 3}}$
&
$(8,6) : \pm \frac{8\,\alpha_{\mp}\,\beta_{\mp}\,\sqrt{J - 2}\,\sqrt{J - 1}\,p\,\sqrt{s}}{\alpha_{\pm}\,\sqrt{J}\,\sqrt{J + 1}\,\sqrt{J + 2}\,\sqrt{J + 3}\,M_{2}}$
\\ \hline
$(8,7) : \frac{8\,\sqrt{J - 2}\,\sqrt{J - 1}\,p\,\left(\pm 3\,\beta_{\pm}\,M_{2} + 2\,\alpha_{\mp}\,p\right)}{\sqrt{J}\,\sqrt{J + 1}\,\sqrt{J + 2}\,\sqrt{J + 3}\,M_{2}}$
&
$(8,8) : \pm \frac{8\,\alpha_{\mp}\,\sqrt{J - 2}\,\sqrt{J - 1}\,p^2}{\sqrt{J}\,\sqrt{J + 1}\,\sqrt{J + 2}\,\sqrt{J + 3}\,\sqrt{s}}$
\\ \hline
\end{tabular}
\caption{Non zero elements of the transformation matrix $\tilde U^{J}_{\pm,\frac{3}{2}\bar{\frac{3}{2}}}$ for
the $\frac{3}{2} \bar{\frac{3}{2}}$ channel.}
\label{tbl:DDtransf}
\end{table*}

Further detailed analyses were performed for the cases where there are
one or two spin three-half particles in the final or initial state. We derived the corresponding
covariant partial-wave amplitudes $T^{J}_{\pm,ij}$ and the projectors ${\cal Y}^{J}_{\pm,ij}$. Since the expressions
are prohibitively tedious they are not presented here. All what is needed in most practical applications are the
transformation matrices from the helicity states to the covariant states. For those explicit results are provided.
We assure that the implied phase-space distributions (\ref{def-rho}) confirm with the desired asymptotic
behavior (\ref{def-rho-constraint}). The matrices for the cases involving one spin one-half and one spin three-half
state  are shown in Tabs. \ref{tbl:DNtransf}- \ref{tbl:NDtransf}. In Appendix B we show as a further example the
projectors for a reaction involving one spin three-half particle only.
For the most tedious case with two spin three-half states, we have derived the transformation matrix in
two steps. First we derive a set of covariant amplitudes that are free of kinematical constraints, even though at
this point they do not have the appropriate asymptotic properties (\ref{def-rho-constraint}). This first step is
defined by the transformation matrix $\tilde U^J_{\pm,\frac{3}{2}\bar{\frac{3}{2}}}$, which is given
in Tab. \ref{tbl:DDtransf}.  The final transformation is
\begin{eqnarray}
U^J_{\pm,\frac{3}{2}\bar{\frac{3}{2}}} = \tilde U^J_{\pm,\frac{3}{2}\bar{\frac{3}{2}}}
 \left(
  \begin{array}{cccccccc}
  \frac{1}{M_{1}\,M_{2}} & 0 & 0 & 0 & 0 & 0 & 0 & 0\\
  \frac{2}{M_{1}\,M_{2}} & \frac{1}{s} & 0 & 0 & 0 & 0 & 0 & 0\\
  0 & 0 & \frac{1}{M_{1}\,M_{2}} & 0 & 0 & 0 & 0 & 0\\
  \pm \frac{2\,J}{M_{1}\,M_{2}} & 0 & \frac{1}{M_{2}\,\sqrt{s}} & \frac{1}{M_{2}\,\sqrt{s}} & 0 & 0 & 0 & 0\\
  0 & 0 & \frac{2\,\left(M_{1}^2 + M_{2}^2 + s\right)}{M_{1}\,M_{2}\,s} & \mp \frac{2}{s} & \frac{1}{s} & 0 & 0 & 0\\
  \frac{-4\,J}{M_{1}\,\sqrt{s}} & 0 & -\frac{2}{M_{1}\,M_{2}} & 0 & -\frac{1}{s} & \frac{1}{M_{1}\,\sqrt{s}} & 0 &       0\\
  \frac{4\,J}{M_{1}\,M_{2}} & \pm \frac{2\,J\,M_{2}}{M_{1}\,s} & \pm \frac{2\,\left(M_{1} \pm M_{2}\right)}{M_{1}\,M_{2}\,\sqrt{s}} & \pm \frac{2}{M_{2}\,\sqrt{s}} & \mp \frac{1}{M_{1}\,\sqrt{s}} & \mp \frac{1}{s} &       \frac{1}{s} & 0\\
  \pm \frac{8\,J\,(M_{1}\mp M_2)}{M_{2}^2\,\sqrt{s}} & 0 & \mp \frac{4}{M_{1}\,M_{2}} & 0 & \pm \frac{6}{s} &       \pm \frac{2}{M_{1}\,\sqrt{s}} & \mp \frac{2}{M_{2}\,\sqrt{s}} & \frac{1}{s}
\end{array}
\right) \,.
\label{defU12}
\end{eqnarray}

\end{widetext}

\section{Conclusions}

We have constructed partial-wave amplitudes for baryon-anti-baryon scattering which are free from kinematical
constraints and frame independent. Those amplitudes are well suited to be used in partial-wave dispersion relations.
Explicit transformations from the conventional helicity states to the covariant states were derived and presented in
this work. We considered  the scattering of two \sphlf, two \spthlf particles, and a \sphlf from a \spthlf particle.

As a necessary intermediate step we identified complete sets of invariant functions that parameterize the scattering
amplitudes of the various processes and are kinematically unconstrained. The latter are expected to satisfy
Mandelstam's dispersion integral representation. Furthermore we constructed a projector algebra that solves
the two-body Bethe-Salpeter scattering equation in the limit of short range forces. It was pointed out that
the existence and smoothness of such a projector basis is closely related to the existence of covariant
partial-wave amplitudes. Explicit expressions for such projectors were presented for the two simplest reactions.

The present paper thus offers an efficient starting point for analyzing
baryon-anti-baryon scattering in a covariant coupled-channel approach that takes into account the constraints set by
micro-causality and coupled-channel unitarity.


\appendix

\section{Expansion in Legendre polynomials}\seclab{parity}

The helicity-partial-wave amplitudes are expressed in terms of Legendre polynomials and the invariant amplitudes.
We introduce a set of auxiliary helicity  amplitudes
\begin{eqnarray}
&& \phi_{ij}^{\pm} = \frac{1}{d^{(J_0)}_{+\lambda, \bar{\lambda}}(\theta) } \,
\tsp{\bar{\lambda}_{1},\bar{\lambda}_{2}}{+\lambda_{1},+\lambda_{2}}
\nonumber \\ && \qquad
 \pm \,\frac{(-1)^{S_{1}-S_{2}}}{d^{(J_0)}_{-\lambda ,\bar{\lambda}}( \theta)}  \,
\tsp{\bar{\lambda}_{1},\bar{\lambda}_{2}}{-\lambda_{1},-\lambda_{2}} \;,
\label{def-helicity}  \\ && \qquad
\quad
 J_{0} = {\rm Min }\left( \left| \bar{\lambda} \right|,\left| \lambda \right|\right) \nonumber \,,
\end{eqnarray}
where $\lambda = \lambda_{1}-\lambda_{2}$ and $\bar{\lambda}
= \bar{\lambda}_{1}-\bar{\lambda}_{2}$ is the total helicity of the initial and final state respectively.
Partial-wave matrix elements can now be expressed in terms of $\phi_{ij}^{\pm}$ and the Legendre polynomials
$P_n(\cos\theta)$. It holds
\begin{eqnarray}
\label{eq:tfromphi}
&&  t^{J}_{\pm,ij} =\langle  \bar \lambda_{1}, \bar \lambda_{2} | \,T_J | \lambda_{1},\lambda_{2} \rangle
\nonumber \\ && \qquad \quad
 \pm  (-1)^{S_{1}-S_{2}}\, \langle  \bar \lambda_{1}, \bar \lambda_{2}| \,T_J | -\lambda_{1},-\lambda_{2} \rangle
\nonumber \\ && \qquad
 \; = \,    \int_{-1}^{1} \frac{d\,\cos\theta}{2} \Big[
        \left(\phi_{ij}^{+} + \phi_{ij}^{-} \right)
            d^{(J_0)}_{\lambda_j,\lambda_i}(\theta)\,
            d^{(J)}_{\lambda_j,\lambda_i}(\theta)
 \nonumber \\ && \qquad \quad
	 \pm   \left(\phi_{ij}^{+} - \phi_{ij}^{-} \right)
            d^{(J_0)}_{-\lambda_j,\lambda_i}(\theta)\,
            d^{(J)}_{-\lambda_j,\lambda_i}(\theta) \Big] \;,
\end{eqnarray}
and
\begin{eqnarray}
&& d^{(J_0)}_{\lambda_j,\lambda_i}(\theta) \,d^{(J)}_{\lambda_j,\lambda_i}(\theta) =
    \sum_{n = \left|J-J_0\right|}^{n = J + J_0} (-1)^{\lambda_j-\lambda_i}\, (2\, n +1)
 \nonumber \\ && \quad \times
        \left(
          \begin{array}{ccc}
            J_0     & J     & n \\
            -\lambda_j & \lambda_j & 0 \\
          \end{array}
        \right)
        \left(
          \begin{array}{ccc}
            J_0     & J     & n \\
            -\lambda_i & \lambda_i & 0 \\
          \end{array}
        \right) P_n(\cos\theta) \;.
\end{eqnarray}

\section{Projectors for the $\frac{1}{2}\bar{\frac{1}{2}} \rightarrow \frac{1}{2} \bar{\frac{1}{2}}$ reaction}

The expressions of the projectors provided in (\ref{result-projectors}) can be further simplified by using
appropriate building blocks, that are derived by suitable derivatives on the function  $Y_J^{(n)}(\bar r, r, s)$. We write
\begin{eqnarray}
&& \partial^\mu  Y_J^{(n)} \equiv \frac{\partial}{\partial\, r^\mu}\,Y_J^{(n)}(\bar r, r, s) =
 (J-n)\,\frac{r^\mu}{r^2} \,Y_J^{(n)}
\nonumber\\
&& \quad +  \left(\frac{\bar{r}^2 \,r^2}{s^2}\right)^{(J-n)/2} P_J^{(n+1)}(x)
\Big[- \frac{\bar{r}^\mu }{\sqrt{{\bar{r}^2 \,r^2}}} +\frac{\bar r \cdot r}{\sqrt{{\bar{r}^2 \,r^2}}} \frac{r^\mu}{r^2} \Big]
\nonumber\\
&& \quad  = -\frac{\bar{r}^\mu}{s} \, Y_J^{(n+1)}
- \frac{r^\mu}{s} \frac{\bar{r}^2}{s} \, Y_{J-1}^{(n+1)}\,,
\nonumber\\
&& \bar \partial^\mu  Y_J^{(n)} \equiv \frac{\partial}{\partial\, \bar r^\mu}\,Y_J^{(n)}(\bar r, r, s) =
\nonumber\\
&& \quad  = -\frac{r^\mu}{s} \, Y_J^{(n+1)}
- \frac{\bar r^\mu}{s} \frac{r^2}{s} \, Y_{J-1}^{(n+1)}\,,
\nonumber\\
&& \bar{\partial}^\mu \partial^\nu Y_J^{(n)} =
- \frac{\bar{r}^\nu }{s}\,\bar{\partial}^\mu Y_J^{(n+1)} - \frac{r^\nu \,\bar{r}^2}{s^2} \, \bar{\partial}^\mu Y_{J-1}^{(n+1)}
\nonumber\\
&& \quad -\,\frac{g^{\mu\nu}}{s} \, Y_J^{(n+1)} - 2\, \frac{\bar{r}^\mu \,r^\nu}{s^2} \, Y_{J-1}^{(n+1)}\,,
\label{partial-derivatives-Y}
\end{eqnarray}
where we used
\begin{eqnarray}
x\, P_J^{(n+1)}(x)= (J-n)\, P_J^{(n)}(x)+ P_{J-1}^{(n+1)}(x)\,.
\end{eqnarray}

The projectors for the $\frac{1}{2}\bar{\frac{1}{2}} \rightarrow \frac{1}{2} \bar{\frac{1}{2}}$ reaction read:
\begin{eqnarray}
&&  {\cal Y}_{\pm,11}^{J} = \mp  \frac{1}{s} \Big( P_\pm \otimes P_\pm \Big) \,Y_J \,,
\nonumber\\
&&  {\cal Y}_{\pm,12}^{J} = \pm  \frac{1}{\sqrt{s}} \,\Big( P_\pm \otimes P_\pm \, \tilde \gamma_{\nu }\Big)
\,\partial^\nu \, Y_J \,,
\nonumber\\
&& {\cal Y}_{\pm,21}^{J} =  \pm \frac{1}{\sqrt{s}} \,\Big( \tilde \gamma_{\mu } P_\pm \otimes \,P_\pm\Big) \,\bar{\partial}^\mu \, Y_J \,,
\nonumber\\
&& {\cal Y}_{\pm,22}^{J}
= \mp \,\Big( \tilde \gamma_{\mu } P_\pm \otimes \, P_\pm \,\tilde \gamma_{\nu }\Big) \,
     \bar{\partial}^\mu \partial^\nu \,Y_J \,,
\end{eqnarray}
where we recall the notation
\begin{eqnarray}
&& \tilde{\gamma}^\mu={\gamma}^\mu- \slashed{w} \,\frac{w^\mu}{s}\,.
\label{def-gammatilde}
\end{eqnarray}

\section{Projectors for the $\frac{1}{2}\bar{\frac{1}{2}} \rightarrow \frac{3}{2} \bar{\frac{1}{2}}$ reaction}

We first provide some useful intermediate results used in the derivation of the projectors
for $\frac{1}{2}\bar{\frac{1}{2}} \rightarrow \frac{3}{2} \bar{\frac{1}{2}}$ reaction. It holds
\begin{eqnarray}
&& 	\,\bar{r}^\alpha \,\langle \tilde \gamma_{\alpha } P_\mp \otimes \,P_\pm \rangle_\mu
  = \frac{ \bar{M}_\mp}{2\,s} \left( s - \bar{M}_{\pm}^2 \right)  \, \langle P_\mp  \otimes  P_\pm  \rangle_\mu \,,
\nonumber \\
&&	r^\alpha\, \langle P_\mp \otimes  P_\pm \, \tilde \gamma_{\alpha }\rangle_\mu =\frac{ M_\pm}{2\,s}
  \left( s - M_\mp^2 \right)  \,\langle P_\mp \otimes  P_\pm \rangle_\mu \,,
\nonumber \\
&&	\langle \tilde \gamma^{\mu } P_\mp \otimes \,P_\pm \rangle_\mu
   = - \frac{ \bar{M}_\pm}{s} \,w^\mu \, \langle P_\mp  \otimes  P_\pm  \rangle_\mu \,,
\nonumber\\
&&  \bar{r}^\mu \,\langle  \bar{\Gamma} \otimes\Gamma \rangle_\mu =
- \frac{s+\bar{M}_+ \bar{M}_-}{2\,s}\, w^\mu \,\langle  \bar{\Gamma} \otimes\Gamma \rangle_\mu \,,
\label{res-appendix}
\end{eqnarray}
where we use the notation of (\ref{def-gammatilde}) and
\begin{eqnarray}
&&\langle  \bar{\Gamma} \otimes\Gamma \rangle_{\mu} =
\bar{u}_\mu (\bar{p}_1)\,\bar{\Gamma}\,v(\bar{p}_2)\,
\bar{v}(p_2)\,\Gamma \,u(p_1)\,.
\end{eqnarray}
The first two on-shell identities in (\ref{res-appendix}) resemble our previous results (\ref{res-44}) used in the derivation of
the $\frac{1}{2}\bar{\frac{1}{2}} \rightarrow \frac{1}{2} \bar{\frac{1}{2}}$ projectors. The last two identities follow from
\begin{eqnarray}
&& \bar u_\mu(\bar p_1) \,\gamma^\mu = 0 = \bar u_\mu(\bar p_1)\,p_1^\mu\,,
\nonumber\\
&&\bar{p}_1^\mu=w^\mu + 2\,\bar{k}^\mu=w^\mu + 2\,\bar{r}^\mu + 2 \,\frac{w\cdot \bar{k}}{s}\, w^\mu \,,
\end{eqnarray}
where we recall (\ref{def-k}, \ref{def-r}). Using the notations of Appendix B we obtain:
\begin{widetext}
\begin{eqnarray}
&&  {\cal Y}_{\pm,11}^{J,\mu} = \pm  \frac{\sqrt{2\, J + 1}}{2 \,s^2}
\Big( \slashed{\bar{p_1}}\,P_\mp \otimes P_\pm \Big)\,w^\mu\, Y_J  \,\,,
\nonumber\\
&&  {\cal Y}_{\pm,12}^{J,\mu} = \mp  \frac{\sqrt{2\, J + 1}}{2 \,s \sqrt{s}} \,\Big(
\slashed{\bar{p}_1} P_\mp \otimes P_\pm \, \tilde \gamma_{\beta }\Big)\, w^\mu \,\partial^\beta \, Y_J \,,
\nonumber\\
&& {\cal Y}_{\pm,21}^{J,\mu} =  \mp \frac{\sqrt{2\, J + 1}}{2\, J\,s \sqrt{s}}
\,\Big( \slashed{\bar{p}_1}\,\tilde \gamma_{\alpha } P_\mp \otimes \,P_\pm\Big)\,w^\mu \,\bar{\partial}^\alpha \, Y_J \,,
\nonumber\\
&& {\cal Y}_{\pm,22}^{J,\mu}
= \mp \frac{\sqrt{2\, J + 1}}{2\, J\,s} \,\Big(
     \slashed{\bar{p}_1} \tilde \gamma_{\alpha } P_\mp \otimes \, P_\pm \,\tilde \gamma_{\beta }\Big)\, w^\mu \,
     \bar{\partial}^\alpha \partial^\beta \,Y_J \,,
\nonumber\\
&&  {\cal Y}_{\pm,31}^{J,\mu}
    = \pm \frac{\sqrt{2\, J + 1}}{ J\,s^2}\,\sqrt{s}\,\Big[
        \Big( P_\mp \otimes P_\pm \Big)\,\Big( s\, \bar{\partial}^\mu  - 2 \, J \, w^\mu \Big)
   + \Big( \slashed{\bar p_1} \, \tilde \gamma_{\alpha } P_\mp \otimes \,P_\pm\Big) \,w^\mu \,\bar{\partial}^\alpha
\Big]\,Y_J\,,
\nonumber\\
&& {\cal Y}_{\pm,32}^{J,\mu}
    = \mp \frac{\sqrt{2\, J + 1}}{ J\,s} \Big[
      \Big( P_\mp \otimes P_\pm \, \tilde \gamma_{\beta }\Big)\,s \, \bar{\partial}^\mu \partial^\beta
\nonumber\\
&& \qquad \qquad
    - \Big(P_\mp \otimes P_\pm \, \tilde \gamma_{\beta } \Big) \,2\,J \, w^\mu \partial^\beta
    +  \Big( \slashed{\bar{p}_1}\,\tilde \gamma_{\alpha }
    P_\mp \otimes \, P_\pm \,\tilde \gamma_{\beta }\Big)\,w^\mu \,\bar{\partial}^\alpha \partial^\beta
      \Big]\,Y_J\,,
\nonumber\\
&& {\cal Y}_{\pm,41}^{J,\mu}
    = \pm \frac{\sqrt{2\, J + 1}}{ J(J-1)\,s^2} \, \Big[
           \Big( \tilde \gamma_{\alpha } P_\mp \otimes \,P_\pm \Big)\,s^2\, \bar{\partial}^\mu \bar{\partial}^\alpha
     -  \Big( \tilde \gamma_{\alpha } P_\mp \otimes \,P_\pm \Big)\, 2\,(J-1)\, s \, w^\mu\,\bar{\partial}^\alpha
\nonumber\\
&& \qquad \qquad
    + \Big( \slashed{\bar p_1} P_\mp \otimes P_\pm\Big)\,4\,J\,(J-1)\,w^\mu
     \Big] \, Y_{J}\,,
\nonumber\\
&& {\cal Y}_{\pm,42}^{J,\mu}
    = \pm \frac{\sqrt{2\, J + 1}}{ J(J-1)\,s\,\sqrt{s}} \Big[
   \Big( \tilde \gamma_{\alpha } P_\mp \otimes \, P_\pm \,\tilde \gamma_{\beta } \Big)\,
   s^2 \, \bar{\partial}^\mu \bar{\partial}^\alpha \partial^\beta
   -\Big(\tilde \gamma_{\alpha } P_\mp \otimes \, P_\pm \,\tilde \gamma_{\beta } \Big)\, 2\,(J-1)\, w^\mu\, s\,
   \bar{\partial}^\alpha \partial^\beta
\nonumber\\
&& \qquad \qquad
 + \Big( \slashed{\bar{p}_1}P_\mp \otimes P_\pm \, \tilde \gamma_{\beta }\Big)\,4\,J\,(J-1)\,w^\mu  \, \partial^\beta
\Big] Y_J \,.
\label{res-projectorsC}
\end{eqnarray}
Note that the last four projectors in (\ref{res-projectorsC}) involve structures already present in the first four projectors.
\end{widetext}


\end{document}